\documentclass[preprint]{aastex}

\usepackage{natbib}

\usepackage{graphicx}

\shorttitle{UV Spectroscopy of Circumnuclear Star Clusters in M83}

\shortauthors{Wofford et al.}

\begin{document}

\title{Ultraviolet Spectroscopy of Circumnuclear Star Clusters in M83}

\author{Aida Wofford, Claus Leitherer}

\affil{Space Telescope Science Institute, 3700 San Martin Drive, Baltimore, MD 21218, USA}

\email{wofford@stsci.edu}

\and

\author{Rupali Chandar}

\affil{University of Toledo, Department of Physics and Astronomy, Toledo, OH 43606, USA}

\begin{abstract}

We analyze archival \textit{HST}/STIS/FUV-MAMA imaging and spectroscopy of 13 compact star clusters within the circumnuclear starburst region of M83, the closest such example. We compare the observed spectra with semi-empirical models, which are based on an empirical library of Galactic O and B stars observed with \textit{IUE}, and with theoretical models, which are based on a new theoretical UV library of hot massive stars computed with WM-Basic. The models were generated with Starburst99 for metallicities of $Z=0.020$ and $Z=0.040$, and for stellar IMFs with upper mass limits of 10, 30, 50, and 100\,M$_\odot$. We estimate the ages and masses of the clusters from the best fit model spectra, and find that the ages derived from the semi-empirical and theoretical models agree within a factor of 1.2 on average. A comparison of the spectroscopic age estimates with values derived from \textit{HST}/WFC3/UVIS multi-band photometry shows a similar level of agreement for all but one cluster. The clusters have a range of ages from about 3 to 20\,Myr, and do not appear to have an age gradient along M83's starburst. Clusters with strong P-Cygni profiles have masses of a few$\,\times10^4$\,M$_\odot$, seem to have formed stars more massive than $30$\,M$_\odot$, and are consistent with a Kroupa IMF from $0.1-100$\,M$_\odot$. Field regions in the starburst  lack P-Cygni profiles and are dominated by B stars. 

\end{abstract}

\keywords{galaxies: individual (NGC 5236) --- galaxies: nuclei --- galaxies: starburst --- galaxies: star clusters --- galaxies: stellar content}

\section{INTRODUCTION}

In the context of disk galaxies, a nuclear or circumnuclear ring is a region of large gas surface density and enhanced star formation, located within 2 kpc of the galactic nucleus. \cite{com10} distinguish between nuclear disks and nuclear rings, by setting the maximum width of the ring to half the ring radius. The most compact rings (ultra-compact nuclear rings, UCNRs) have radii smaller than 200 pc \citep{com08}. The Atlas of Images of Nuclear Rings \citep{com10} is the most complete atlas of galactic nuclear rings to date and includes 101 barred and unbarred disk galaxies with rings. The dust nuclear rings around elliptical and S0 galaxies constitute a different class of galactic ring with no associated star formation. The present work is not concerned with the latter rings. Circumnuclear rings in disk galaxies give rise to starbursts \citep{ser65, mao96, bon98}, which observationally are ring- or arc-shaped. Note that the term ``starburst'' does not have a general definition \citep{con08, kna09}. A ``circumnuclear starburst'' is a region composed of star condensations with individual masses ranging from a few\,$\times10^4$ to greater than $10^6$\,M$_\odot$, bright in the ultraviolet (UV) because they have recently emerged from their birth clouds (the nuclear galactic ring), and contain hot and massive O and/or B stars.

Proposed scenarios for the origin of star-forming galactic rings are galaxy collisions or mergers, accretion of intergalactic gas, and resonances caused by the actions of rotating bars or other non-axisymmetric disturbances on the motions of disk gas clouds \citep{but96}.  Note that in their analysis of H$\alpha$ and continuum images of 327 local disk galaxies, \cite{kna09} found no significant increase in the central concentration of star formation as a result of the presence of a close companion. On the other hand, there is strong observational evidence for a connection between bars and circumnuclear starbursts \citep{hun99}. In addition, non-starburst barred galaxies may eventually become starburst barred galaxies \citep{jog05}. The location of nuclear rings in barred galaxies may be set by inner Lindblad resonances (ILRs), which can prevent gas from flowing further in. Such resonances can cause gas to accumulate in circumnuclear rings where clusters can form \citep{but96}. Alternative models for the location of starbursts in barred galaxies can be found in \cite{shl90} and \cite{reg97}, while \cite{van09} have a model for the migration of circumnuclear star clusters and nuclear rings.

Circumnuclear starbursts in barred galaxies are perhaps the most numerous class of nearby starburst regions \citep{mao01}. Their detailed study can provide a stepping stone for studies of starbursts at higher redshifts. Indeed, local starburst galaxies (including circumnuclear starbursts), show star formation rate densities \citep{meu97}, UV colors \citep{meu99, ade00}, and spectral morphologies \citep{con96, pet00}, similar to those of high redshift Lyman-break galaxies (also see \citealt{hec05}). 

Determining the ages, masses, and stellar initial mass functions (IMFs) of individual star clusters within circumnuclear starbursts is important for understanding the relations between (1) galaxy dynamics and interactions and the origin of circumnuclear star clusters; (2) massive star feedback and the fueling of active galactic nuclei (AGN, \citealt{che09, hao09}); and (3) young massive star clusters and globular clusters \citep{dig09, elm10}. In this study, we determine the ages and masses, and constrain the IMFs of 13 star clusters in Messier 83 (M83, NGC 5236), which hosts the nearest example of a circumnuclear starburst in a barred galaxy.

M83 is a nearby (4.5$\pm$0.3\,Mpc, \citealt{thi03}), southern (Dec[J2000]\,$=-29^\circ51'56''$), nearly face-on ($i=24^\circ$, \citealt{rog74}), SAB(s)c grand-design spiral galaxy \citep{deV92}, with an optical disk spanning $12'.9\times11.5'$ ($17.0\,$kpc$\,\times15.2\,$kpc). \cite{wik04} have estimated that at the distance of M83, a central black hole of moderate size and luminosity would be detectable. However, the galaxy shows no indication from radio or X-ray observations of hosting an AGN \citep{cow85, sor02}. M83 has a close dynamical companion in the dwarf irregular galaxy NGC 5253 \citep{rog74}, which contains a starburst nucleus of $40-50\,$pc in size (\citealt{cal97}, \citealt{cal99}, \citealt{tre01}). 

M83's arc-shaped circumnuclear starburst has been observed at wavelengths ranging from the radio to the X-rays (see \citealt{dop10} for references). It spans about 200\,pc in length and 35\,pc in thickness, and it is located between two rings of dusty gaseous material which may be coincident with two inner Lindblad resonances \citep{elm98}. The starburst qualifies as an UCNR (ring radius of $<200\,$ pc), and as shown in this work, it breaks up into about 20 compact FUV bright star clusters. Two proposed scenarios for its origin are, the interaction of M83 with its neighbor NGC 5253 about  1\,Gyr ago \citep{van80}, and/or the merging of M83 with a dwarf satellite in the past. The former scenario is supported by the "finger" of H I gas projecting outward from the halo of M83 toward NGC 5253 \citep{rog74}, and by the presence of off-disk H I clouds in M83 \citep{mil09}. The merger scenario is based on kinematic studies using R band \citep{mas06}, near-infrared \citep{tha00,dia06}, and submillimeter \citep{sak04} spectroscopy, which have revealed the presence of a controversial second nucleus, hidden by dust, more massive than the optical nucleus, and located $4"\pm 1"$ northwest of the latter. Recent N-body simulations by \cite{rod09}, which account for this second nucleus, predict the disruption of the circumnuclear starburst in less than an orbital time. However \cite{hou08} have concluded that there is no second nucleus in M83, since they cannot confirm the local maximum in the stellar velocity dispersion at the location of the ``second nucleus'' found by \cite{tha00}. A third scenario for the origin of the starburst could be internal galaxy dynamics associated with the bar. 

The ages of circumnuclear star clusters in M83 have been previously estimated. Based on the analysis of CO absorption and Br$\gamma$ emission-line data \cite{pux97} found that the northwestern end of M83's starburst is the youngest, while the southeastern end is the oldest. \cite{har01} compared \textit{HST}/WFPC2 photometry with theoretical population synthesis models and found that most of M83's circumnuclear star clusters are less than 10 Myr old, with clusters younger than 5 Myr preferentially located along the northwestern end. The statistical analysis of the photometry of \cite{har01} by \cite{dia06} confirms the age gradient along the starburst. In their analysis, \cite{dia06} divided the $100^\circ$ starburst region into 25$^\circ$ angular sectors, and considered the mean cluster age as well as the oldest age within each sector, finding 5\,Myr for the oldest cluster in the northwest and 25\,Myr for the oldest cluster in the southeast. The above age gradient suggests that star formation occurs in an ordered manner, and supports the idea that the starburst is fed by the inflow of bar-driven material. If real, this gradient has profound implications for the formation and evolution of circumnuclear starbursts. For this reason, in this work, we age-date 13 clusters within the northern portion of M83's starburst, using three different techniques, i.e., by comparing archival \textit{HST}/STIS FUV spectroscopy of the clusters with semi-empirical and new fully theoretical model spectra, and by deriving the ages from high spatial resolution \textit{HST}/WFC3 multi-band optical photometry.  Although our study does not cover the full extent of the starburst's arc, this does not affect our results. In addition, we analyze the massive star population in regions between the clusters.

Another goal of this work is to constrain the upper IMFs of our clusters. This is particularly interesting in the metal rich circumnuclear environment of M83. M83's circumnuclear H II regions have an oxygen to hydrogen number density ratio of 12+log(O/H)=9.1$\pm$0.2 \citep{bre02}, and a corresponding fraction of all metals to all elements by mass (hereafter, metallicity) that is intermediate between Z=0.020 and Z=0.040, and supersolar (12+log[O/H]$_\odot=\,$8.69$\pm$0.05, Z$_\odot$=0.013, \citealt{asp09}). We estimate the cluster masses from spectroscopy and photometry, and check which clusters have formed stars more massive than $\,>30\,$M$_\odot$. The random sampling of the stellar IMF may prevent this from happening in low mass clusters \citep{wei10}.

This paper is organized as follows: \S~\ref{observations} describes the observations and data reduction, \S~\ref{models} presents the models used in the estimate of the spectroscopic ages and masses of the clusters, \S~\ref{analysis} explains our procedure for determining the spectroscopic and photometric properties of the clusters, in \S~\ref{discussion} we present and discuss our results, finally, we conclude in \S~\ref{conclusion}.

\section{OBSERVATIONS AND DATA REDUCTION}\label{observations}

We analyzed archival FUV imaging and spectroscopy of compact star clusters and field regions located between about $5''$ ($=$110\,pc) and $9''$ ($=$198\,pc) of M83's optical center, and spanning an arc of $\sim$200\,pc in size. The data were taken with the STIS onboard the \textit{HST} as part of programs GTO-7577 and GTO-8785 (PI: Sara Heap). We complemented these data with optical images of the same clusters taken with \textit{HST}'s WFC3 as part of program GO/DD-11360 (PI: O'Connell). \cite{cha10} discuss properties derived from optical photometry of star clusters within the larger field of view covered by the WFC3 images. Our main focus is on the FUV data, but we also briefly discuss the optical data.

Figure~\ref{fig_composite} shows a false-color logarithmic-scale image of M83's arc-shaped circumnuclear starburst. The image is a composite of the STIS image taken with the F25QTZ (FUV) filter, and WFC3 images taken with the F657N (H$\alpha$) and F555W (V) filters. North is at the top and east is to the left. The optical nucleus of M83 is the large yellow-green dot in the upper left. The starburst starts northwest of the nucleus, wraps about the nucleus towards the southeast, and is punctuated by about 20 UV bright clusters. The FUV bright clusters appear white, while fainter ones appear blue. The location of SN1968L, which was identified by \cite{dop10}, is indicated. At M83's distance, 1'' represents about 22 pc. The two dusty rings discovered by \cite{elm98} and embracing the starburst are clearly seen. The clusters with available STIS FUV spectroscopy are labeled. Note that although clusters 3, 4 and 7 seem to be composed of several bright knots clustered together, one knot always dominates the brightness. Figure~\ref{fig_composite} was kindly produced for us by W. P. Blair.

\subsection{FUV Imaging and Spectroscopy with STIS}\label{fuvobs}

The FUV data set consists of a direct image and four two-dimensional (2-D) spectrograms. The direct image is projected onto the MAMA detector, which has a field of view of 25''x25'' and a plate scale of 0$^".$024 pixel$^{-1}$. The spectrograms, one long slit (25''x2'') and three multi-object (slitless), were taken with the G140L grating, which has an average dispersion of 0.584 \AA~pixel$^{-1}$. Table~\ref{tab_data} gives the observation date (col. 2), exposure time (col. 3), aperture position angle in degrees East of North (col. 4), filter name (col. 5), wavelength coverage for the detector$\,+\,$filter combination (col. 6), and extension of the most processed data file available from the Hubble Data Archive of the FUV observations.

The data were partially processed through the On-the-Fly Reprocessing (OTFR) pipeline of the archive, at retrieval time. This included data linearization, dark subtraction, flat fielding, and summing of the repeated observations. In addition, the pipeline corrected the direct image for geometric distortion. After processing through the pipeline, in order to increase the  signal-to-noise (S/N), we added together exposures taken with the same instrument configuration, i.e., the two slitless exposures taken with the F25SFR2 aperture, which we refer to as the F25SFR2 exposure.

Because the spectrograms were taken in either a slitless or a wide slit mode with the G140L grating, the spectral traces have large offsets in the dispersion direction. Therefore, the OTFR pipeline did not wavelength calibrate or convert these spectra to absolute flux. In order to extract wavelength-calibrated one-dimensional (1-D) spectra from the 2-D spectra, we first located the position of each spectral trace along the cross dispersion direction, and then used the X1D module (McGrath, Busko, \& Hodge, STIS ISR 99-03) available in the Image Reduction and Analysis Facility (IRAF, \citealt{tod86,tod93}) for this task. Several iterations were required in order to find the optimal value of xoffset, which is the parameter used by X1D for shifting a spectrum along the dispersion direction. We used the absorption feature C II $\lambda\lambda1334,1335+\lambda1336$ (hereafter C II 1335), for finding the optimal value of xoffset. In our spectra, this feature is unresolved and is a blend of Milky Way interstellar and M83 stellar components.

For the width of the extraction box we used the value recommended in the STIS manual for point sources, i.e., 11 pixels \citep{bro02}. Background spectra extracted from outer portions of the 2-D spectrograms were assumed to represent purely geocoronal emission (air glow) from H I Lyman-$\alpha$ (Ly$\alpha$) at $1216\,$\AA~and/or,  [O I] $1302+1305+1306\,$\AA, depending on the aperture, and were subtracted from the cluster spectra. We did not subtract the underlying diffuse light because of its large variations as a function of position.

\begin{figure}
\plotone{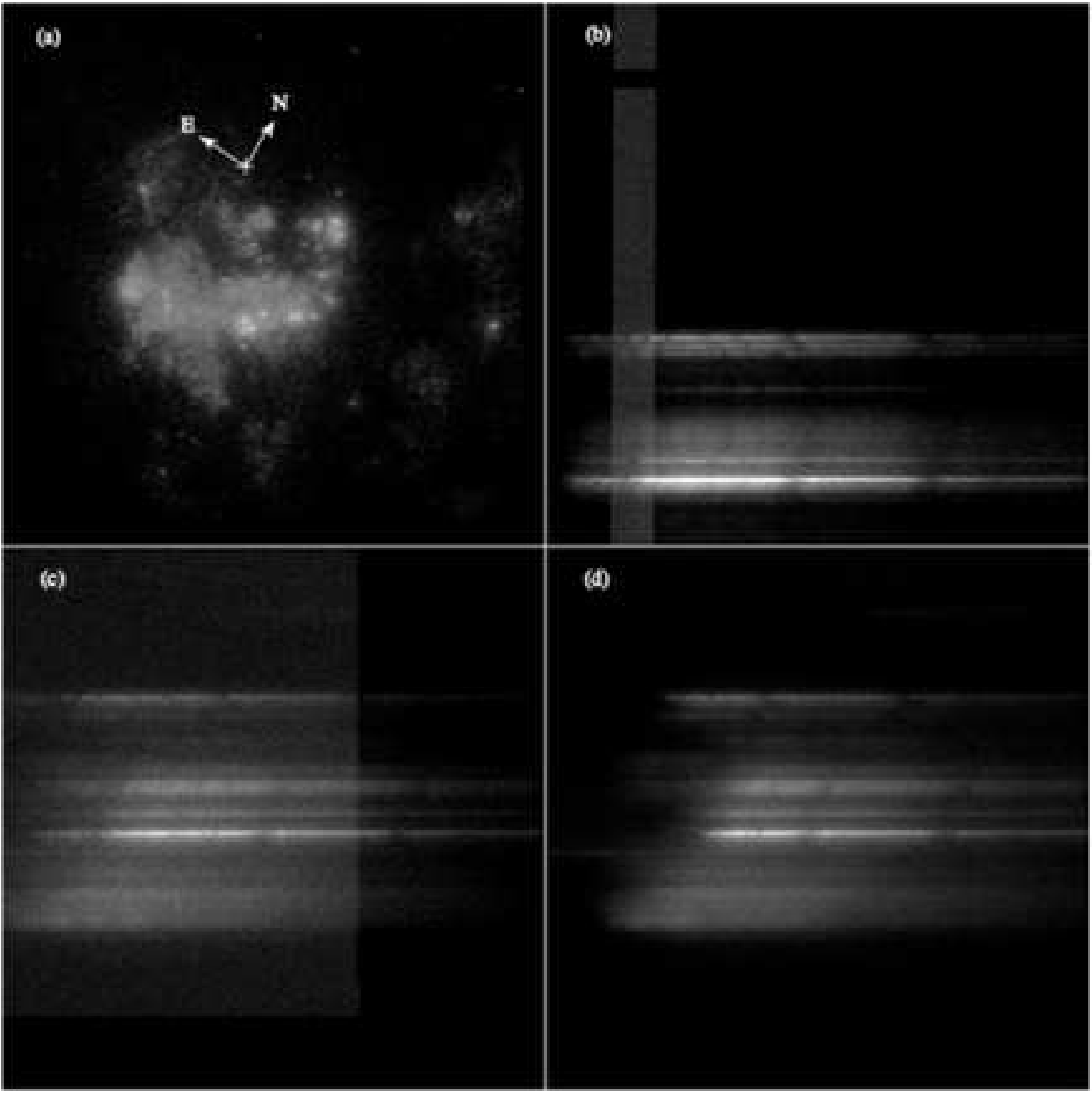}
\caption{\textit{HST}/STIS/FUV-MAMA data set. Panel (a). 25''x25'' logarithmic-scale image ($1475-1700\,$\AA~band), with M83's optical center (cross), compass directions, and spatial scale overplotted. Panel (b). 2-D $1150-1700\,$\AA~25''x2'' long-slit spectrogram. The vertical band is due to Ly$\alpha$ air glow filling the slit. Note the horizontal shift of the spectral traces due to the use of a wide slit. Panel (c). 2-D $1150-1700\,$\AA~slitless spectrogram. The Ly$\alpha$ air glow contaminates much of the image. Note the horizontal shift of the spectral traces due to the use of no slit. Panel (d). 2-D $1275-1700\,$\AA~slitless  spectrogram. Note the lack of Ly$\alpha$ air glow contamination.}
\label{fig_fuvset}
\end{figure}

The direct image and the 2-D spectrograms from the STIS are shown in the four panels of Fig.~\ref{fig_fuvset}. Panel (a) features the 25''x25'' direct image in logarithmic scale, with the location of M83's optical center marked with a cross, and the spatial scale marked with a vector. Panels (b) and (c) show the long-slit and the multi-object spectrograms respectively, both taken with the same aperture (25MAMA). The vertical band in panel (b) is due to Ly$\alpha$ airglow filling the 2''-wide slit (it is seen more clearly in the electronic version). The Ly$\alpha$ airglow contaminates much of the slitless spectrogram shown in panel (c). Finally, panel (d) features the sum of the two slitless spectrograms taken with the F25SFR2 aperture, which rejects the Ly$\alpha$ airglow. No contamination from the [O I] air glow is apparent in panel (d) because the emission brightness geocoronal [O I] is typically less than $10\%$ of that of geocoronal Ly$\alpha$, depending on the time of observation and the position of the target relative to the Earth \citep{mis03}. 

We used the slitless exposures for extracting spectra of those clusters falling outside the slit, as well as for studying the effect of rejecting the Ly$\alpha$ air glow on our results.

The spectral resolution, depends on the spatial extent of the star clusters. In order to estimate the spectral resolution of our spectra, we determined the full width at half maximum (FWHM) of the cross dispersion profiles of the spectral traces. For our point-like representative, GD 153, the FWHM is 0.09''. Due to blending with the profiles of close neighbours, we were only able to determine the FWHM of four clusters by our method. For clusters 1, 4, 10, and 11+12 we obtained FWHM of 0.17'', 0.30'', 0.19'', and 0.23'', respectively, which suggests that the clusters are partially resolved. Since these are neither point sources nor fully extended sources, we expected the spectral resolution of our long-slit spectra to be intermediate between the values given in the STIS instrument handbook, i.e., 0.9\AA~(point source) and 48\AA~(extended source). Multi-component Gaussian fitting of the C II feature at $\lambda\lambda1334,1335+\lambda1336$, carried out with the Peak Analysis code (available at ftp://ftp.ncnr.nist.gov/pub/staff/dimeo/pan.zip) reveals that the FWHM of this feature ranges from 2.5-7~\AA.  Our field spectra are expected to have the resolution of a fully extended source. 

We corrected the cluster spectra for reddening due to the presence of dust in the Galaxy and in M83. For the Galaxy's contribution to the reddening, we adopt $E(B-V)=0.06$ \citep{sch98}, and we assume the Galactic extinction curve of \cite{fit99}. For M83's contribution, we (i) fitted a power law of the form $F\propto\lambda^\beta$ to the continuum in the range $1240-1600\,$\AA, (ii) assumed that any deviation of $\beta$ from $-2.6$, which is the expected value for a dust free starburst, is due to extinction by local dust, and (iii) dereddened the observed spectrum until $-2.6$ was reached, using the starburst obscuration law of \cite{cal00}.

\subsection{Matching the Clusters with their Spectral Traces}\label{match}

\begin{figure}
\plotone{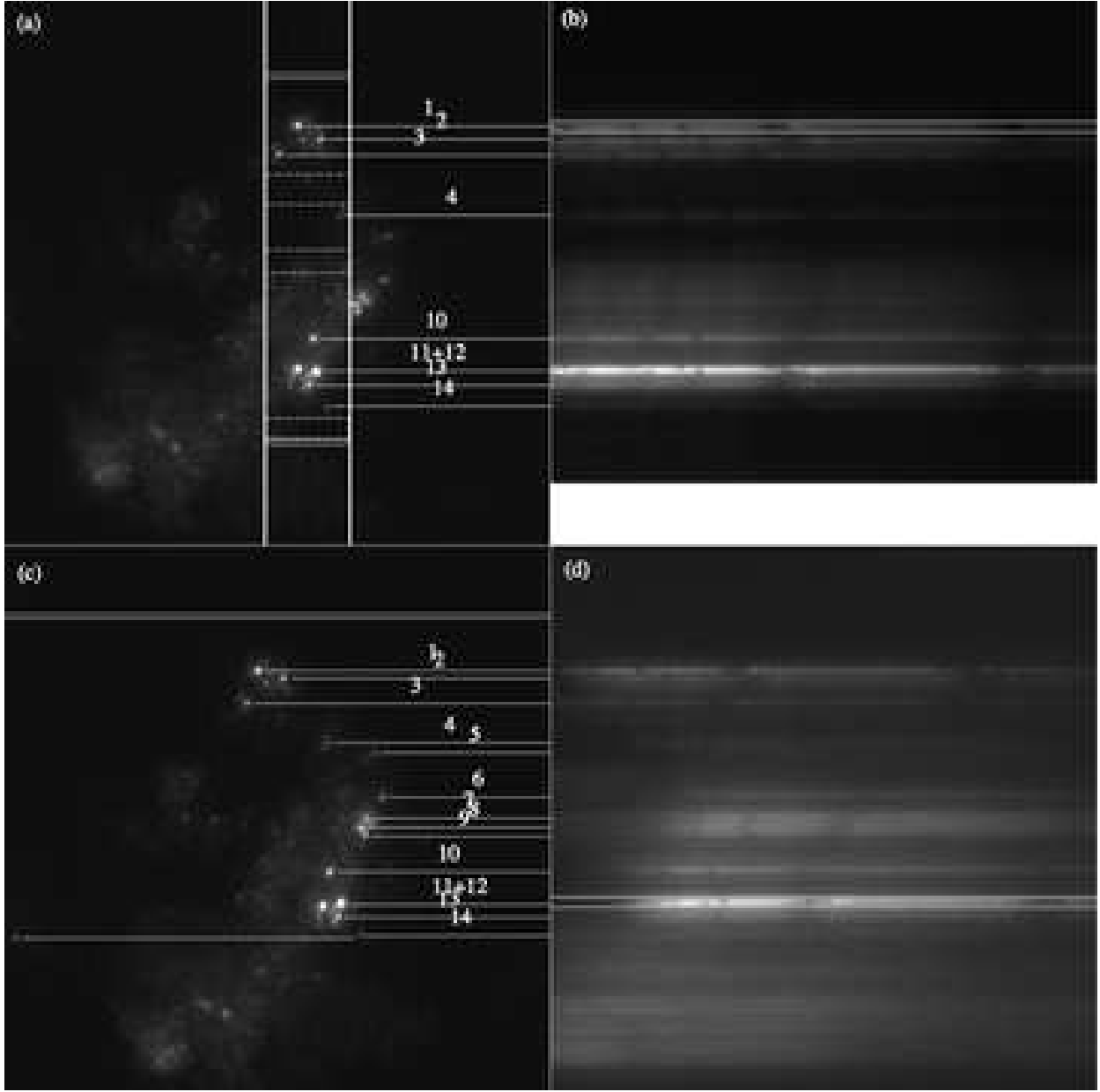}
\caption{Enlarged portions of the STIS images arranged to show the spatial correspondence between star clusters and spectral traces. Panels (a) and (c). Image rotated by 42.5$^\circ$ and 57.9$^\circ$ East of North, respectively. Lines joining the clusters and their spectral traces are overplotted in the two panels. The 2''-wide slit (thick vertical lines) and the extraction regions used for the background and the field spectra (solid and dashed horizontal lines within slit, respectively) are overplotted in panel (a). Only the top background region is shown in panel (c) since the bottom one falls outside of the figure. Panels (b) and (d). 2-D $1150-1700\,$\AA~long-slit  and $1275-1700\,$\AA~slitless spectrograms. The size of the box used for extracting the 1-D spectra is shown. }  
\label{fig_fuvmatch}
\end{figure}

In order to match each cluster with its FUV spectral trace, we started by directly comparing the positions of the clusters in the STIS direct image, with the positions of the spectral traces in the 2-D spectrograms. For this, we rotated the image and vertically shifted the spectrograms until the orientations matched, as shown in Fig.~\ref{fig_fuvmatch}, which shows enlarged portions of the STIS image, side by side with two spectrograms. Note that the slitless spectrograms all have the same orientation, while the orientation of the long-slit spectrogram is different. Therefore, in Fig.~\ref{fig_fuvmatch}, we show only the slitless spectrogram taken with the F25SFR2 aperture. In panel (a) of Fig.~\ref{fig_fuvmatch}, we indicate the position of the long-slit, the locations where we extracted the sky and field region spectra, and the correspondence between clusters within the slit and their spectral traces. Panel (c), which shows the same image as panel (a), but with a slightly different orientation, shows only one sky region due to the fact that the second falls outside of the field displayed in the panel.

Unfortunately, and as can be seen in Fig.~\ref{fig_fuvmatch}, in spite of the different orientations of the spectrograms, the spectral traces of the two brightest clusters, 11 and 12, always overlap. Furthermore, as seen in panel (d), the faint spectral traces of clusters 4 and 5 also overlap, as do the traces of cluster 14 and a source not in our sample. Fortunately, we can use the long-slit spectra for clusters 4 and 14. However, cluster 5 was rejected from our analysis due to its poor spectral trace.

\begin{figure}
\plotone{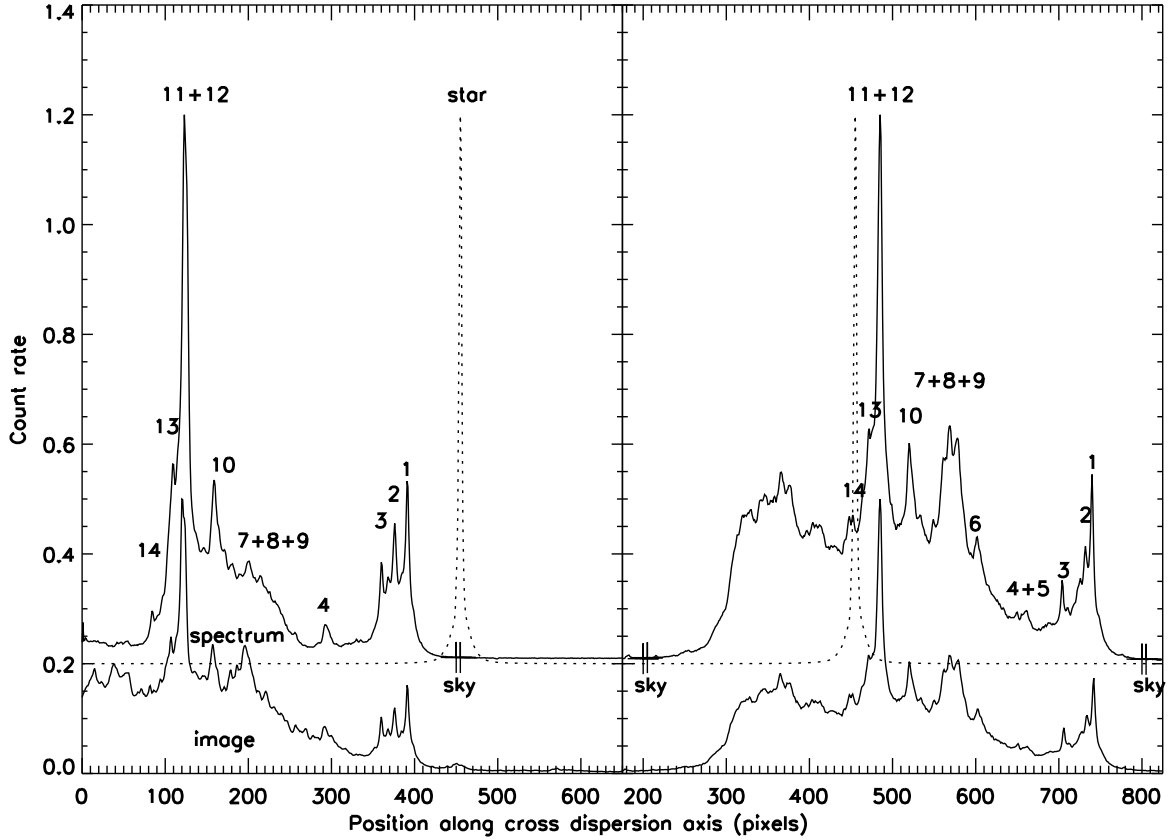}
\caption{Plots showing the spatial correspondence between the star clusters in the STIS image, and the long-slit and slitless spectral traces (left and right panels, respectively). In each panel, the top curve represents the cross dispersion profiles of the spectral traces, while the bottom curve represents the radial profiles of the clusters. The clusters in our sample are labeled 1-14. The background subtraction regions are labeled ``sky''. The dotted curve is the profile of a point-like white dwarf observed with the same configuration that was used for the long-slit spectrum. The profiles were obtained and scaled as explained in the text.}  
\label{fig_fuvprof}
\end{figure}

The spatial correspondence between star clusters and their spectral traces can be checked in Fig.~\ref{fig_fuvprof}, where we compare the cluster radial profiles derived from the STIS direct image (bottom curves in left and right panels), with the cross dispersion profiles derived from the long-slit (top curve, left panel) and the slitless (top curve, right panel) 25MAMA spectrograms. Also shown in Fig.~\ref{fig_fuvprof} is the cross-dispersion profile of a point-like source, represented by white dwarf GD 153 (top dotted curve in left and right panels), observed as part of program 7097, in the same configuration as the long-slit spectrogram. For convenience, we divided each curve in Fig.~\ref{fig_fuvprof} by the amplitude of the tallest peak, which for the clusters is the amplitude of target 11$+$12 (which overlap), and for the white dwarf is its own amplitude. For clarity, the top curves in Fig.~\ref{fig_fuvprof} (cross dispersion profiles) were all shifted by 0.2. The clusters and regions used for the background subtraction are labeled. 

For the radial profiles, we first rotated the image to match the orientation of each spectrogram, and we then obtained the count rate at a given vertical position by integrating over the horizontal direction. The radial profiles in the left and right panels look slightly different because of the different angles by which we rotated the image. We note that although the peaks for clusters in our sample are dominated by emission from the clusters themselves, diffuse light and fainter sources also contribute to their shape and amplitude. This however, does not affect the identification of the spectrum of each cluster. 

For the spectral profiles, the count rate at a given position in the cross dispersion direction was obtained by integrating over the dispersion axis. Note the higher background of the top right curve with respect to the top left curve, which is due to contamination with Ly$\alpha$ air glow over a larger portion of the detector in the slitless spectrogram (right panel of Fig. ~\ref{fig_fuvprof}).  

\subsection{Optical Imaging with WFC3}\label{vis}

Circular aperture photometry was performed on the clusters using the IRAF task PHOT, using an aperture radius of 3~pixels and a background annulus with radii of 10 and 13 pixels. Aperture corrections, based on the estimated size of each cluster, were applied to convert the fixed aperture magnitudes to total magnitudes. The instrumental magnitudes in the F336W, F438W, F555W, F658N, and F814W filters were converted to the VEGAMAG photometric system by applying zeropoints available at the following URL: http://www.stsci.edu/hst/wfc3/phot\_zp\_lbn. We refer to the broad-band filters as the ``$U$'', ``$B$'', ``$V$'', and ``$I$'' magnitudes for simplicity, although they were not transformed to the Johnson-Cousins system. For more details, refer to \cite{cha10}.

\section{MODEL SPECTRA}\label{models}

We compared our observed reddening corrected FUV spectra with synthetic dust free spectra generated with the Windows version \citep{lei09} of the widely used package Starburst99 (S99, \citealt{lei99,vaz05}). The main input parameters in S99 are 1) the star-formation law, 2) the IMF, 3) the metallicity and stellar evolution tracks, and 4) the stellar spectral library.

We tried the following parameters. 1) A single burst of star formation, since \cite{bre02} ruled out the possibility that star formation within individual clusters proceeded continuously. Indeed, in M83's circumnuclear region, the He I 5876 absorption line is detected in H II regions ionized by the youngest clusters, but not in those ionized by clusters older than about $6-7$\,Myr. If star formation was continuous, the line strength ratio of He I 5876 to H$\beta$ would be the equilibrium value in the older clusters. 2) IMFs with high mass limits of 10, 30, 50, and 100\,$M_\odot$. 3) The stellar evolution tracks for non-rotating stars of \cite{sch92} for masses $M<12\,M_\odot$, and of \cite{mey94} for masses $M\ge12\,M_\odot$, at metallicities $Z=0.020$ and $Z=0.040$. Note that from H II region measurements, \cite{bre02}  found that the metallicity of M83's nuclear region is $Z\simeq0.030$, which is in between the above two metallicities. 4) An empirical stellar library of Galactic O and B stars, as well as a new UV high resolution theoretical library for massive stars of metallicities $Z=0.013$, and $Z=0.030$. The difference in metallicity between the stellar evolutionary tracks and the theoretical stellar library is due to the use of different solar compositions. Indeed, the WM-Basic models use the revised solar composition of Asplund et al. (2005), which has a lesser content in metals than the solar composition of the stellar evolution tracks. Note that there are no empirical UV stellar libraries for $Z>0.020$. 

We describe the empirical and the new fully theoretical stellar spectral libraries in \S~\ref{libraries}. We discuss the advantages and disadvantages of using the theoretical library in \S~\ref{theo_lib}. Hereafter, we refer to single stellar population (SSP) spectra generated with the empirical stellar library as the semi-empirical models, and to SSP spectra computed with the theoretical stellar library, as the theoretical models. We compare theoretical models at $Z=0.020$ and $Z=0.040$ in \S~\ref{solar_vs_supersolar}. We compare theoretical models corresponding to different high mass limits of the IMF in \S~\ref{upimf}. Finally, we compare the two sets of models in \S~\ref{old_vs_new}.

\subsection{Stellar Spectral Libraries}\label{libraries}

The semi-empirical models were built from spectral libraries of Galactic O and B stars observed in the $1000-1200\,$\AA~range at 0.12\,\AA~resolution with the Far Ultraviolet Spectroscopic Explorer \citep{pel02}, and in the $1200-1800\,$\AA~range at 0.75\,\AA~resolution with the International Ultraviolet Explorer \citep{rob93, deM00}. Comparisons of the semi-empirical models with observations of local and distant galaxies generally yield quite good agreement \citep{leii10}.

The theoretical models were built from a spectral library of hot massive stars ($T_\textrm{eff}\gtrsim20,000\,$K, $M\gtrsim8\,M_\odot$) that was computed as described in Leitherer et al. (2010), using the stellar atmosphere code WM-Basic \citep{pau01}, following the approach pioneered by \cite{rix04}. The models span the spectral range $900-3000\,$\AA~at high (0.5\,\AA) resolution. This resolution is sufficient to resolve the main UV stellar wind lines, which have widths in excess of 1000\,km\,s$^{-1}$.

\subsection{New Theoretical Spectral Library}\label{theo_lib}

The strengths and weaknesses of the theoretical stellar library are the following. 1) The two main publicly available codes optimized for modeling the UV spectra of OB stars with winds are CMFGEN \citep{hil98} and WM-Basic. WM-Basic, which we utilized, solves the radiative transfer and the hydrodynamics self-consistently to derive the density structure of the stellar wind, while CMFGEN assumes an analytical approximation for the wind structure a priori. 2) Wolf-Rayet (W-R) stars, whose stellar evolution is quite uncertain, are modeled using WM-Basic atmospheres. This ensures that their continuum fluxes are accounted for, but W-R-specific lines are not reproduced correctly. Fortunately, W-R stars in the local universe do not contribute significantly to the UV spectrum of OB stellar populations, except for He II $\lambda$1640, which should be viewed with care. In individual W-R stars, the equivalent width (EW) of the He II 1640 emission component ranges from 3 to 100~\AA \citep{con90}, with a preference for smaller values (3-30~\AA or so) for late-type WN stars, which tend to dominate observed statistics due to their high absolute visual magnitude. In starburst clusters, the typical EW of the 1640 bump is about 2-3\,\AA~\citep{cha04}. If EW=15\,\AA~is the typical value for late-type WN stars, this suggests a continuum diluting about 80\% of the light at 1640~\AA, and W-R stars cannot contribute more than 20\% to the UV continuum. The dilution is caused by the strong continuum of O main sequence and O supergiant stars. 3) We omit the geometrical effect of clumps in the stellar wind, but the associated X-ray emission from shocks propagating in the wind is included\footnote{WM-Basic assumes a random distribution of shocks in the wind, where the hot shocked gas is collisionally ionized and excited and emits spontaneously into and through an ambient “cool” stellar wind. This basic model was further refined by \cite{fel97} who accounted for the post-shock cooling zones of radiative and adiabatic shocks. This approach has been implemented in WM-Basic.}.  Clumping affects the lines from trace ions, e.g., O VI $\lambda$1035 and O V $\lambda$1371 \citep{bou05}. In addition, the strength of the wind lines depends on the assumed ionization balance.  While photo-ionization is the primary mechanism determining the ionization state of the wind, X-rays arising from shocked material can be important as well (e.g., \citealt{pau94}).  4) We omit Stark broadening of the hydrogen lowest Lyman lines, which is computationally expensive. A pure Doppler profile of hydrogen is very similar to one with Stark broadening included, as long as the upper principal quantum of the transition is small \citep{rep05}. This is the result of the Stark width's dependence on the fourth power of the upper principal quantum number. In practice, this means Stark broadening becomes negligible in hydrogen lines at higher energies and originating at lower densities. Lyman-$\alpha$, owing to its large optical depth, is formed far out in the wind even at relatively low mass-loss rates. Therefore, Stark broadening is not important at ages younger than 10 Myr. 5) The evolution of massive stars in binary systems and of rotating massive stars requires further study before it can be included in the models \citep{vaz07}. 6) The models can be used for interpreting the rest-frame UV spectra of objects with metallicities other than near-solar or those of the Large ($Z=0.007$) and Small ($Z=0.002$) Magellanic Clouds \citep{mae99}.

\subsection{Theoretical Spectra at $\mathbf{Z=0.020}$ and $\mathbf{Z=0.040}$}\label{solar_vs_supersolar}

Because there are no empirical UV spectral stellar libraries at $Z>0.020$, we have to rely on fully theoretical star cluster spectra, at $Z=0.040$. Here we discuss differences in the theoretical spectra of single stellar populations that were computed with stellar evolution tracks corresponding to metallicities of $Z=0.020$ and $Z=0.040$.

\begin{figure}
\plotone{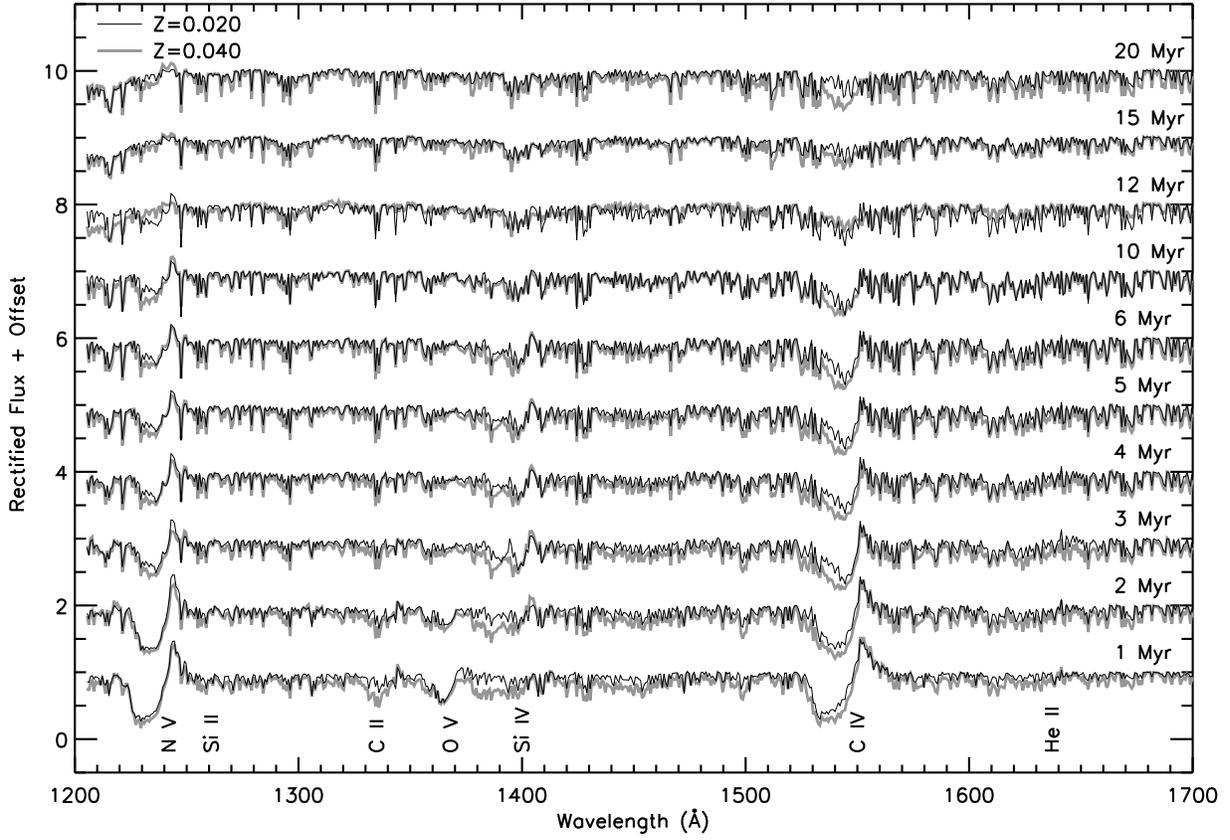}
\caption{Comparison of SSP theoretical spectra with $Z=0.020$ (thin black curves) and $Z=0.040$ (thick grey curves, red in electronic version). A Kroupa IMF from 0.1 to 100\,$M_\odot$ was used. Each spectrum represents an age step from 1 to 20 Myr, as labeled on the right. The spectral features discussed in the text are labeled at the bottom.}
\label{fig_solar_vs_supersolar}
\end{figure}

Line-driven wind theory predicts lower mass-loss rates and lower wind terminal velocities for O stars of lower metallicity \citep{cas75, pul96}, and therefore, weaker wind features in the spectra of these stars. Observations of MS O stars in the Magellanic Clouds and the Galaxy support the theory. For instance, see \cite{mok07}, where empirical stellar wind results for the Milky Way, the LMC and the SMC are compared with the OB star theoretical wind models of \cite{vin01}, which are an extension of line driven wind theory. WM-Basic models do a good job of reproducing lower wind terminal velocities at lower metallicity \cite{lei10}, as can be seen in Fig.~\ref{fig_solar_vs_supersolar}, which shows a comparison between the $Z=0.020$ and the $Z=0.040$ models, for ages from one to 20 Myr (as labeled on the right), in the wavelength range $1200-1700\,$\AA, at 0.75\,\AA~resolution. For instance, compare the absorption components of  C IV 1550 at $Z=0.020$ and $Z=0.040$. In addition, the photospheric lines are deeper at the higher metallicity, as expected. 

\subsection{Model Spectra with Different IMFs}\label{upimf}

\begin{figure}
\epsscale{0.8}
\plotone{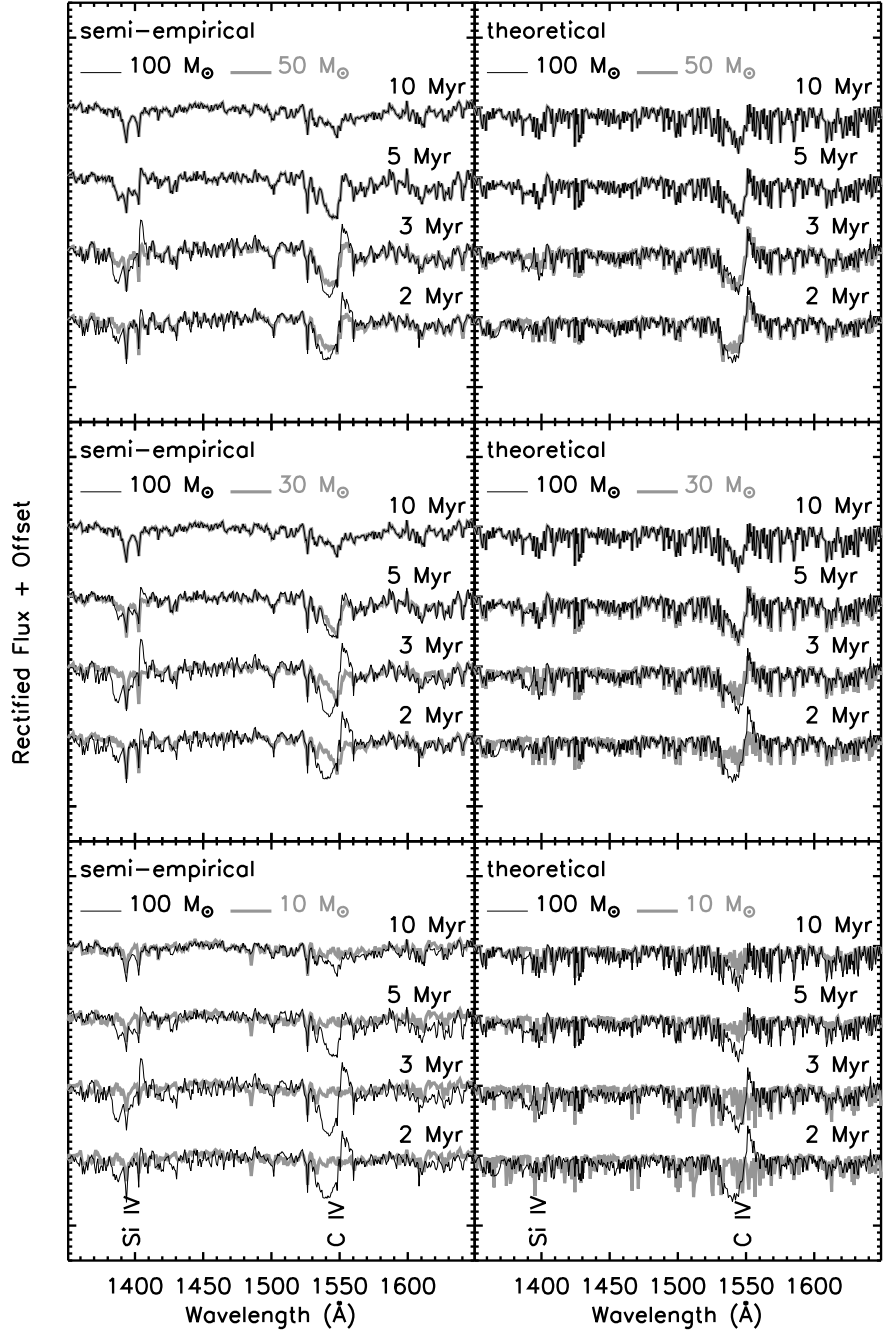}
\caption{Comparison of SSP spectra with high mass limits of 10, 30, 50, and 100\,$M_\odot$ for the IMF. The thin black curves correspond to a Kroupa IMF from 0.1 to 100\,$M_\odot$, while the thick grey curves (red in electronic version) correspond to the high mass limits indicated in each panel. Semi-empirical and theoretical models are shown on the left and right panels, respectively. $Z=0.020$. The region $1350-1650\,$\AA~is shown at ages 2, 3, 5, and 1\, Myr, as labeled on the right. The age sensitive features Si IV 1400 and C IV 1550 are labeled in the bottom two panels.}
\label{fig_upimf}
\end{figure}

Figure~\ref{fig_upimf} shows semi-empirical and theoretical models for SSPs with different high mass limits of the stellar IMF. The figures show ages 2, 3, 5, and 10\,Myr, and focus on C IV 1550  and Si IV 1400, whose profiles are sensitive to the distributions of O main sequence and O supergiant stars, respectively. All models in Fig.~\ref{fig_upimf} have a lower mass limit of $0.1\,$M$_\odot$. The thin black curves use a Kroupa IMF from 0.1 to 100\,M$_\odot$, while the thick grey curves( red in the electronic version) use IMF high mass limits ($M_{up}$) of 10, 30, and 50\,M$_\odot$.

Lowering $M_{up}$, i.e., gradually removing the O and the most luminous B stars, has the following effects. (i) As $M_{up}$ decreases, the P-Cygni profiles of Si IV 1400 and of C IV 1550 become weaker. In particular, the emission component of these profiles is non-existent at $M_{up}=30\,M_\odot$ for the semi-empirical models, and drops in strength more gradually in the theoretical models. Also the strong absorption component of these profiles is purely photospheric at $M_u=10\,M_\odot$, for both models. (ii) For the theoretical models at $M_{up}=10\,M_\odot$, the absorption lines are dominated by photospheric lines from B stars.  

\subsection{Semi-Empirical versus Theoretical Spectra}\label{old_vs_new}

\begin{figure}
\plotone{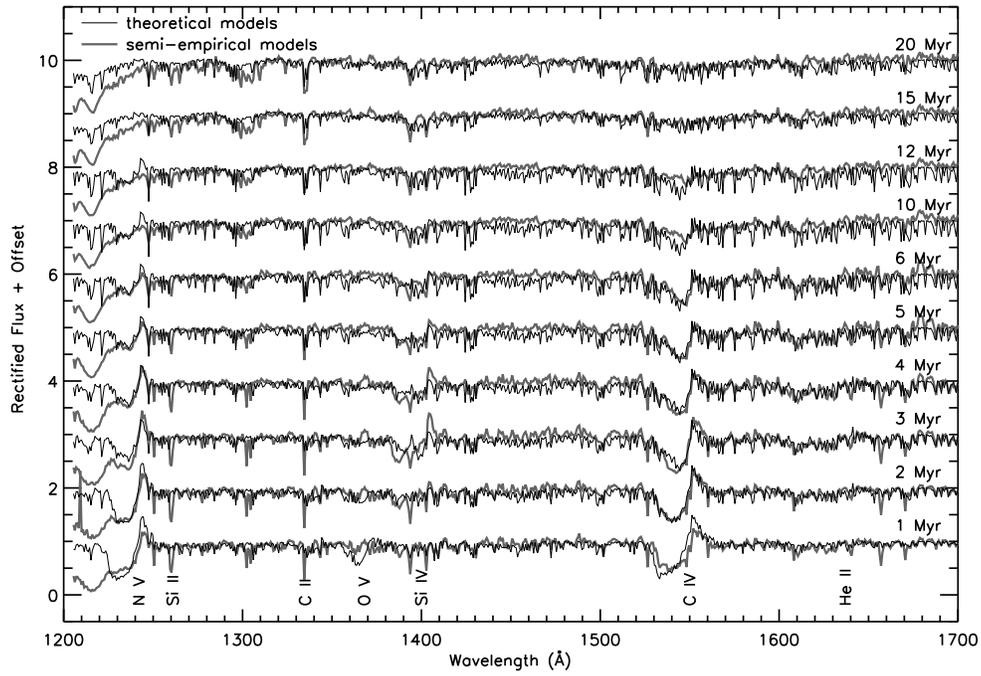}
\caption{Comparison of SSP spectra generated with theoretical (thin black curves) and empirical (thick grey curves, red in electronic version) stellar libraries. $Z=0.020$. A Kroupa IMF from 0.1 to 100\,$M_\odot$ was used. Each spectrum represents an age step from 1 to 20 Myr, as labeled on the right. The spectral features discussed in the text are labeled at the bottom.}
\label{fig_old_vs_new}
\end{figure}

Here we discuss differences between the semi-empirical and the theoretical models at $Z=0.020$. Figure~\ref{fig_old_vs_new} shows the theoretical and semi-empirical models, using the same format of Fig.~\ref{fig_solar_vs_supersolar}. There are several noteworthy features. 1) In the region below 1240 \AA~the semi-empirical spectra suffer from Ly$\alpha$ absorption in the Galactic plane. 2) The P-Cygni profiles of the resonant doublets of N V at 1238.8+1242.8 \AA~(hereafter N V 1240) and of C IV at 1548.2+1550.8 \AA~(hereafter, C IV 1550), which originate in the stellar wind, decrease in strength as the O stars evolve and disappear. 3) The resonant doublet of Si IV at 1393.8+1402.8 \AA~(hereafter, Si IV 1400) is absent in Main Sequence (MS) O stars, and develops P-Cygni profiles in giant and supergiant O stars. This is why the P-Cygni profile of Si IV 1400 is strong at ages $3-5\,$Myr, when O supergiants are present. Note that the lack of strong N V 1240, C IV 1550, or Si IV 1400 P-Cygni profiles, in the spectrum of a star cluster, indicates either that the cluster did not form very massive O stars, or that the cluster is too old to show the signatures of such stars. 4) At 3 Myr, the semi-empirical model has somewhat stronger N V, Si IV, and C IV lines than the theoretical model. At this particular time step, stars with stronger winds make a stronger contribution to the semi-empirical than to the theoretical spectrum. This difference serves as a reminder of a crucial assumption made when linking model spectra to stellar evolution models. In the case of the theoretical spectra, one simply adopts the effective temperature of the atmospheres for matching the positions of the evolutionary tracks. In contrast, for the empirical spectra, we obtain the effective temperatures from spectral types via a temperature scale. Starburst99 assumes the relation between spectral type and effective temperature of O stars derived by \cite{mar05}. We note that switching between the relations of Martins et al. and the earlier work of \cite{vac96} or \cite{sch82} modifies the peak of the Si IV emission by almost 10\%. Careful comparisons with observations should provide guidance for choosing the preferred approach in the synthesis models. In addition, the WM-Basic models use the revised solar composition of Asplund et al. (2005), which has a lesser content in metals than the solar composition of the stellar evolution tracks. While the revised solar abundances have a small effect on the evolutionary tracks of massive stars, they do affect the computed UV spectra via changed opacities and wind properties. 5) Many of the narrow absorption lines in the semi-empirical models are of interstellar origin. For the first four Myr, the absorption line Si II $\lambda$1264 is stronger in the semi-empirical models because of the interstellar contribution. 6) O V $\lambda$1371, which is strong at ages $\lesssim2\,$Myr, i.e., in the hottest stars, is stronger in the theoretical models due to the neglect of wind clumping \citep{bou05}.
 
\section{ANALYSIS}\label{analysis}

\subsection{Spectroscopic Ages and Masses}\label{ana_spec}

Table~\ref{tab_observed} allows an assessment of the uncertainties in the reddened flux at 1500\,\AA, $F_{1500}$, and in the power law index of the reddened FUV continuum, $\beta_{\textrm{obs}}$, for the 13 clusters in our final sample. For each spectral exposure and each cluster, columns (2)-(12) in the table give (1) the cluster ID; (2) and (3) the right ascension and declination measured from the STIS direct image; (4)-(6) the median signal to noise (S/N) of the reddened spectrum; (7)-(9) $F_{1500}$; and (10)-(12) $\beta_{\textrm{obs}}$.

The faint clusters 4 and 14 have the lowest S/N values. The rest of the clusters have reasonably good S/N ($\ge10$). The $F_{1500}$ fluxes from the slitless exposures taken with the 25MAMA and F25SFR2 apertures agree within $10\%$. Therefore, the Ly$\alpha$ airglow has little effect on $F_{1500}$. The slit blocks the light from nearby sources and lowers $F_{1500}$ by $20-50\%$. The value of $\beta_{\textrm{obs}}$ (i) increases by $20\%$ when the wavelength range for fitting the continuum is extended (compare the two values in column 10); (ii) varies by at most $10\%$, due to geocoronal Ly$\alpha$ contamination (compare columns 11 and 12); and (iii) increases by at most $10\%$ when the slit is removed (compare columns 10 and 11 or 12). The properties of clusters $1-3$ derived from the slitless spectra are less reliable than those derived from the long-slit spectra, and are only shown for comparison.

Since the two slitless exposures yield similar values in Tab.~\ref{tab_observed}, we decided to discard the exposure spanning the shortest wavelength range, i.e., the one taken with the F25SFR2 aperture. For the nine clusters within the slit, i.e., 1, 2, 3, 4, 10, 11, 12, 13, and 14, the `good' data span the wavelength range $1240-1690\,$\AA, while for the rest of clusters, i.e., 6, 7, 8, and 9, they span the range $1240-1670\,$\AA.
 
We adopted the algorithm developed by \cite{tre01} for deriving the age and mass of each cluster from the best fit model to the data. The algorithm takes as input the de-reddened cluster spectrum and a grid of Starburst99 models spanning ages from 1 to 20\,Myr in step sizes of 0.1\,Myr. The goodness of the fit is characterized by $\chi^2$, where $\chi^2=(o_i-m_i)^2w_i/\sigma_i^2$, and where $o_i$ represents the observed data for the $i$th pixel, $m_i$ the model data, $\sigma_i$ the error in the observed spectrum, and $w_i$ the assigned weight. The weighting scheme eliminates from consideration the interstellar lines, and gives the stellar wind lines a weight 1.5 times greater than that of the continuum. Prior to fitting, the resolution of the models was degraded to match the approximate resolution of the observed spectra. The spectroscopic masses were derived by (i) fitting power laws to the continua of the observed and the model spectra of each cluster, where the model corresponds to the best fit age and a stellar mass of 10$^6\,$M$_\odot$; (ii) computing the ratio of the observed to the model power law at each wavelength point; (iii) taking the average of the sum of the ratios obtained in (ii); and (iv) multiplying the result by 10$^6\,$M$_\odot$.

The metallicity of M83's starburst is $Z\simeq0.030$ with a large error bar (see \citealt{bre02}). We considered theoretical models with metallicities of $Z=0.020$ and $Z=0.040$. 

\subsection{Photometric Ages and Masses} 

\cite{cha10} derived the age and extinction for each cluster by finding the minimum $\chi^2$ when comparing observed UBVIH$\alpha$ magnitudes with those predicted by the Charlot \& Bruzual (2009, private communication) stellar population synthesis models, assuming a Milky Way-type extinction curve \citep{fit99}, metallicities of $Z=0.017$ and $Z=0.034$, and a Chabrier IMF from $0.1-100$\,M$_\odot$ \citep{chab03}. Here, the narrow band filter contains both stellar continuum and nebular line emission, i.e. no continuum subtraction was performed for the narrow band image, and the predictions were modeled as line plus continuum.

The mass of each cluster was estimated from its extinction-corrected $V$ band luminosity and the age-dependent mass-to-light ratio from the stellar population models, assuming a Chabrier stellar IMF and a distance of 4.5~Mpc. The ages and masses derived using this technique are compared to the spectroscopic values below.

\section{RESULTS}\label{discussion}

\subsection{Metallicity and Theoretical Spectra}

On average, the theoretical spectra fit the data better for $Z=0.040$ than for $Z=0.020$ (on average, the reduced $\chi^2$ is 1.3 times larger at $Z=0.020$ than at $Z=0.040$, with a standard deviation of 0.3). Note that some of the differences between the data and the best-fit models are not due to the metallicity choice, but to the quality of the data, which suffers from crowding, as well as from worse spectral resolution than when a narrow slit is used. In any case, on average, the spectroscopic ages and masses at the two metallicities are identical. Hereafter, we only consider spectroscopic results at $Z=0.020$, which is the metallicity of the semi-empirical models.  

\subsection{IMF}

\begin{figure}
\epsscale{0.8}
\plotone{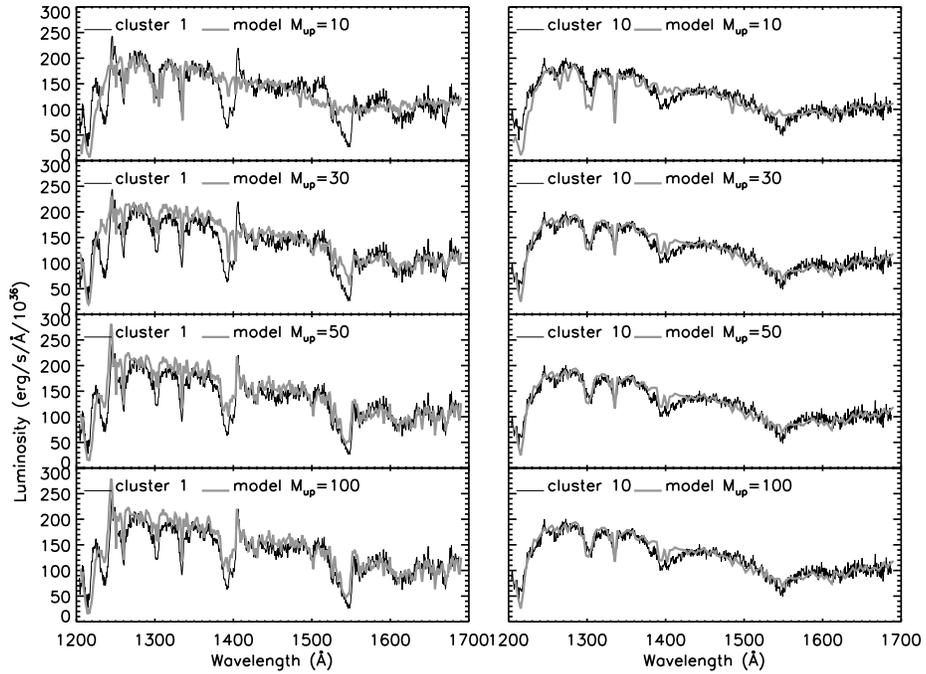}
\caption{Observed spectra of clusters 1 and 10 (thin black curves) versus semi-empirical models corresponding to IMF upper mass limits of $M_{up}=$\,10, 30, 50, and 100\,M$_\odot$ (thick grey curves, red in electronic version).}
\label{fig_1and10}
\end{figure}

Cluster 14 is the youngest ($\sim$3 Myr, average of spectroscopic and photometric ages) and the least massive (photometric mass of about $4\times10^3$\,M$_\odot$) in our sample.   According to \citealt{wei10} cluster 14 should have a maximum stellar mass of about 80 M$_\odot$ (i.e. below 100 M$_\odot$).  Unfortunately, the spectrum of this cluster is too noisy to reliably pin down its maximum stellar mass.  Cluster 1 is the next best cluster in our sample to test the upper end of the stellar IMF, and has a mean age of $\sim$4\,Myr and a photometric mass of  $\sim$3$\times10^4$\,M$_\odot$. As shown in Fig.~\ref{fig_1and10}, left panel,  the strong Si IV 1400 and C IV 1550 P-Cygni profiles of cluster 1 cannot be reproduced with models having maximum stellar masses of 30\,M$_\odot$ or less.  Models with maximum stellar masses of 50 and 100\,M$_\odot$ fit cluster 1 equally well. Note that \cite{bre02} found optical signatures of the presence of WN stars in the region of our clusters $1-3$. The P-Cygni profile of He II at 1640~\AA, which is characteristic of WN stars, is barely detected in the FUV spectra of clusters  $1-3$, but given the point that we make in \S~\ref{theo_lib}, this is not surprising. The rest of clusters with strong P-Cygni profiles of N V, Si IV and C IV are consistent with having formed stars more massive than 30\,M$_\odot$. The right panel of Fig.~\ref{fig_1and10} shows that the upper mass limit of the IMF is harder to constrain for cluster 10, which is without P-Cygni profiles. Clusters with strong absorption profiles of Si IV and C IV are consistent with having formed at least some B stars, if we assume that the absorptions are not of interstellar origin. 

\begin{figure}
\epsscale{0.8}
\plotone{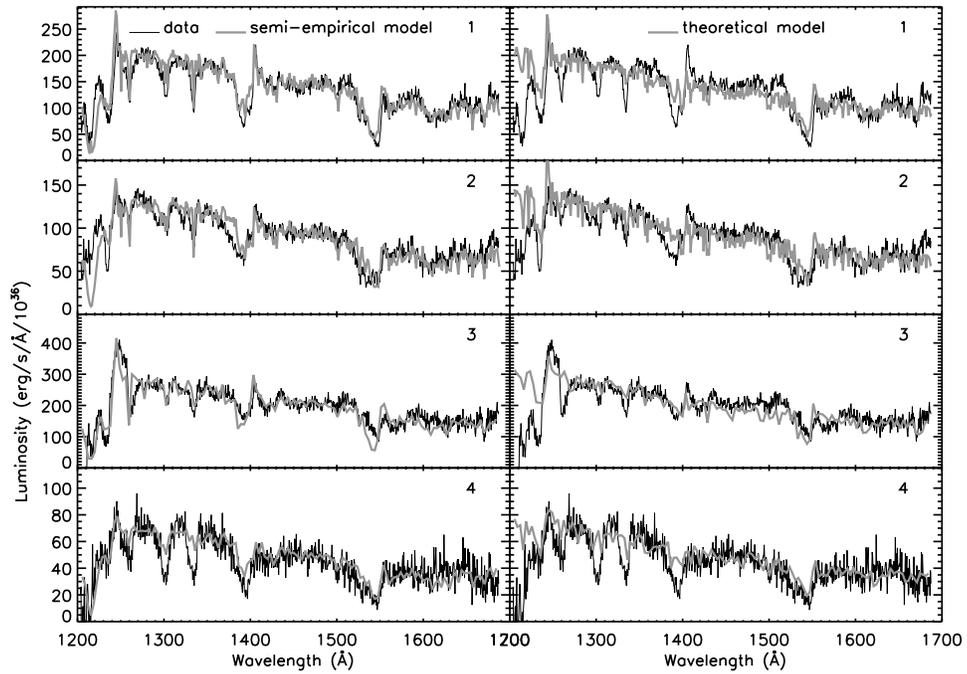}
\caption{Background and reddening corrected rest-frame spectra of clusters $1-4$ (thin black curves), with best fit models overplotted (thick grey curves, red in electronic version). The left panels show the semi-empirical models while the right panels show the theoretical models. The y-axis gives the luminosity as a function of wavelength from $\sim$1200-1700\,\AA.}
\label{fig_fits_1-6}
\end{figure}

\begin{figure}
\epsscale{0.8}
\plotone{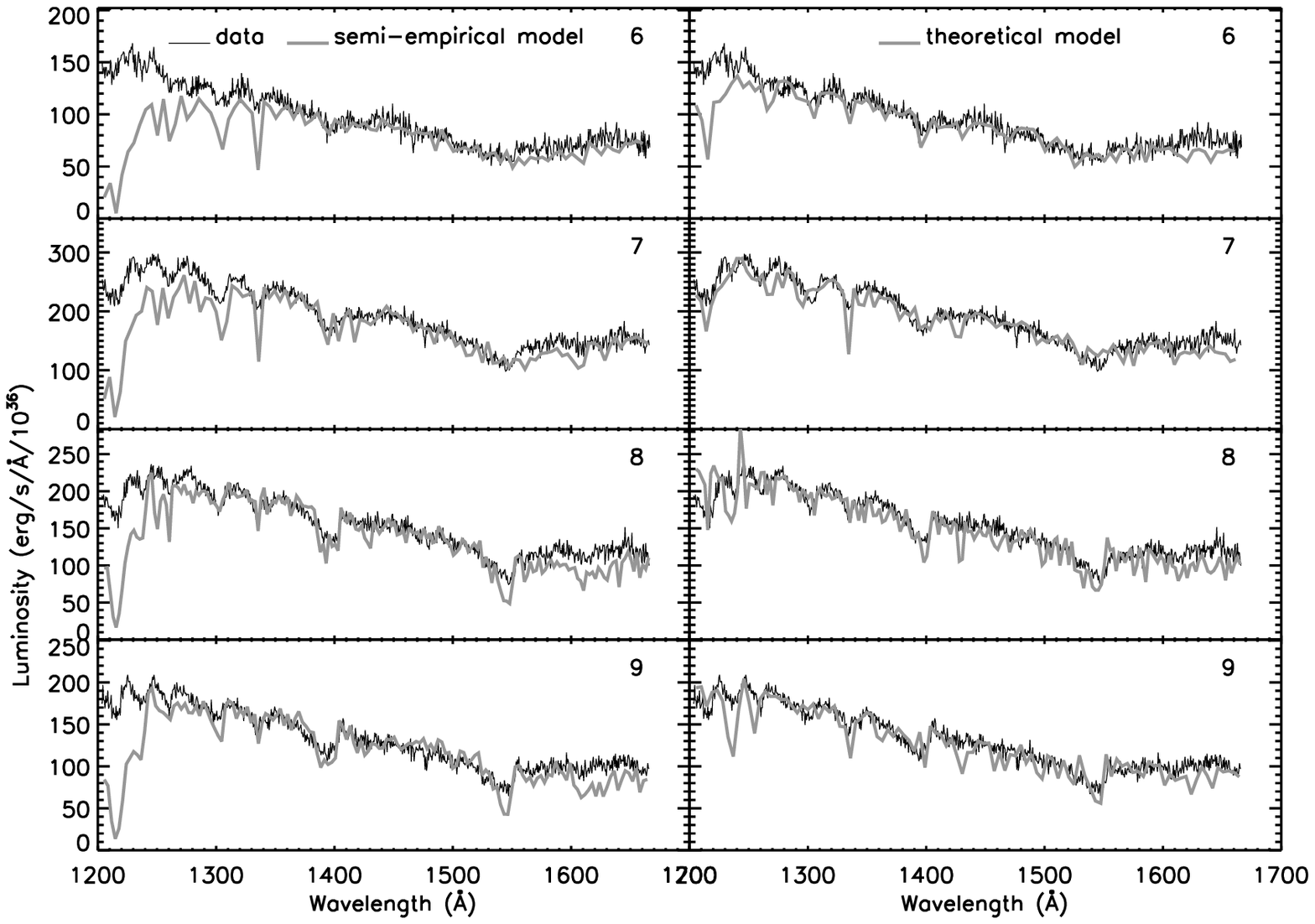}
\caption{Same as Fig.~\ref{fig_fits_1-6} but for clusters $6-9$.}
\label{fig_fits_7-11}
\end{figure}

\begin{figure}
\epsscale{0.8}
\plotone{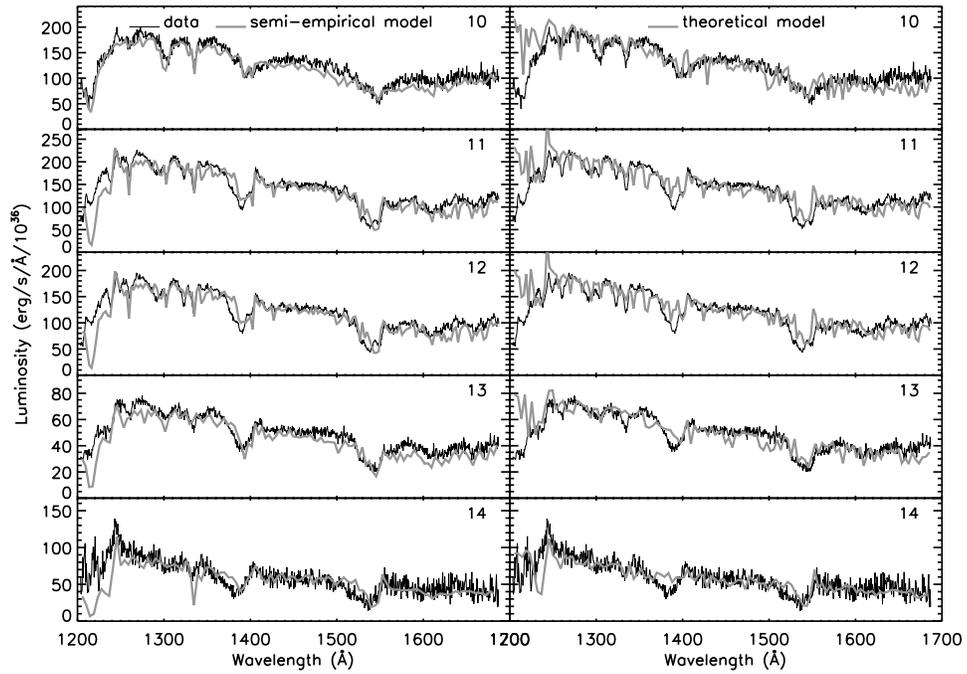}
\caption{Same as Fig.~\ref{fig_fits_1-6} but for clusters $10-14$.}
\label{fig_fits_12-14}
\end{figure}

Figures~\ref{fig_fits_1-6} to~\ref{fig_fits_12-14} show the background and reddening corrected rest-frame spectra of the clusters in our sample, along with best fit semi-empirical and theoretical models corresponding to a metallicity of $Z=0.020$ and a Kroupa IMF from $0.1-100$\,M$_\odot$. We used the latter models to determine the spectroscopic ages and masses of all clusters, including those without strong P-Cygni profiles of N V, Si IV, and C IV in their spectra. Therefore, we are assuming that clusters without strong P-Cygni profiles formed massive O stars but are older than $\sim$12 Myr. Another possibility for the latter clusters is that they are younger than $\sim$12 Myr and did not form very massive O stars.

\subsection{Spectroscopic versus Photometric Age}

Here, we compare our ages derived using spectroscopic and photometric techniques. We also compare our ages with estimates from the literature.

\begin{figure}
\epsscale{1.2}
\plottwo{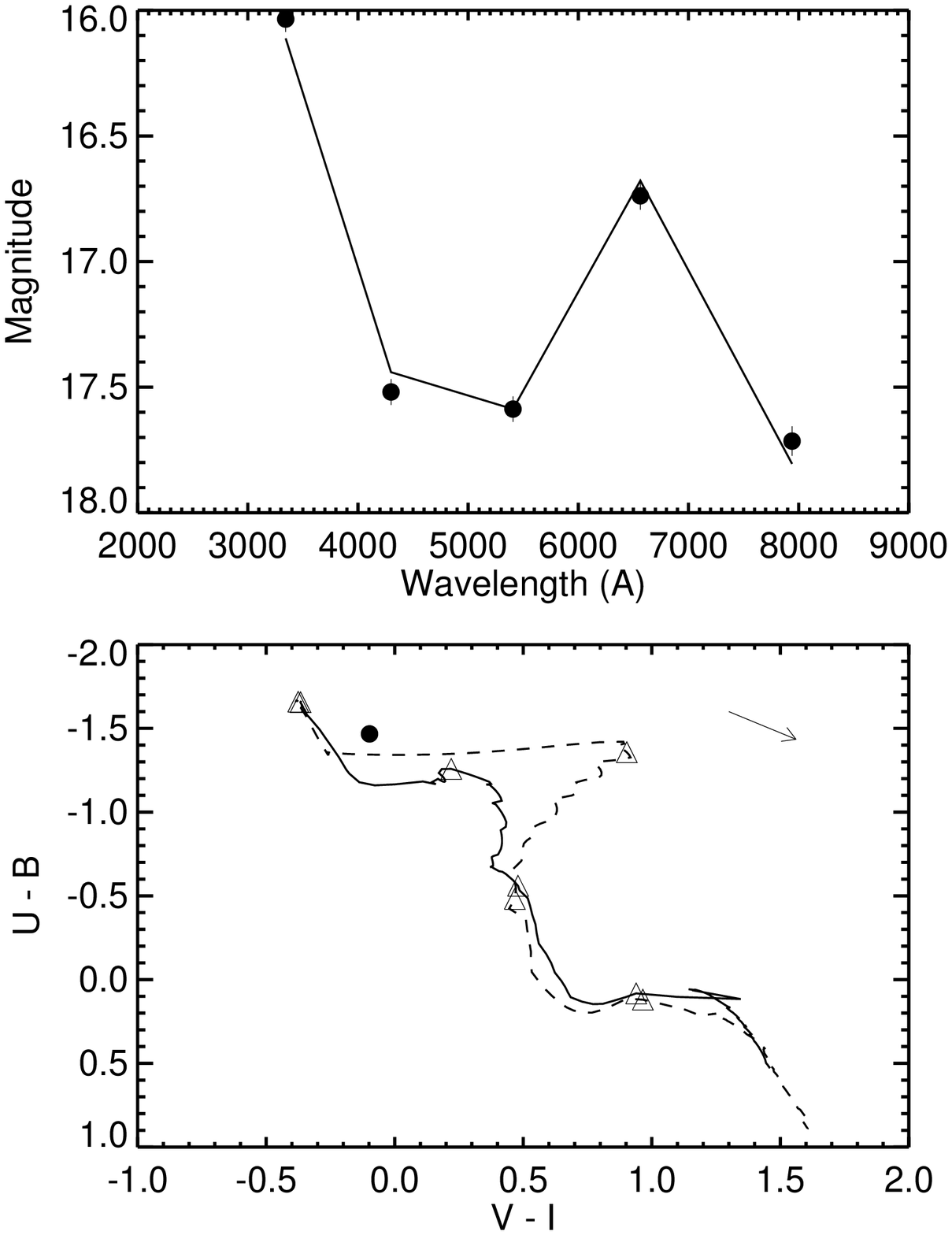}{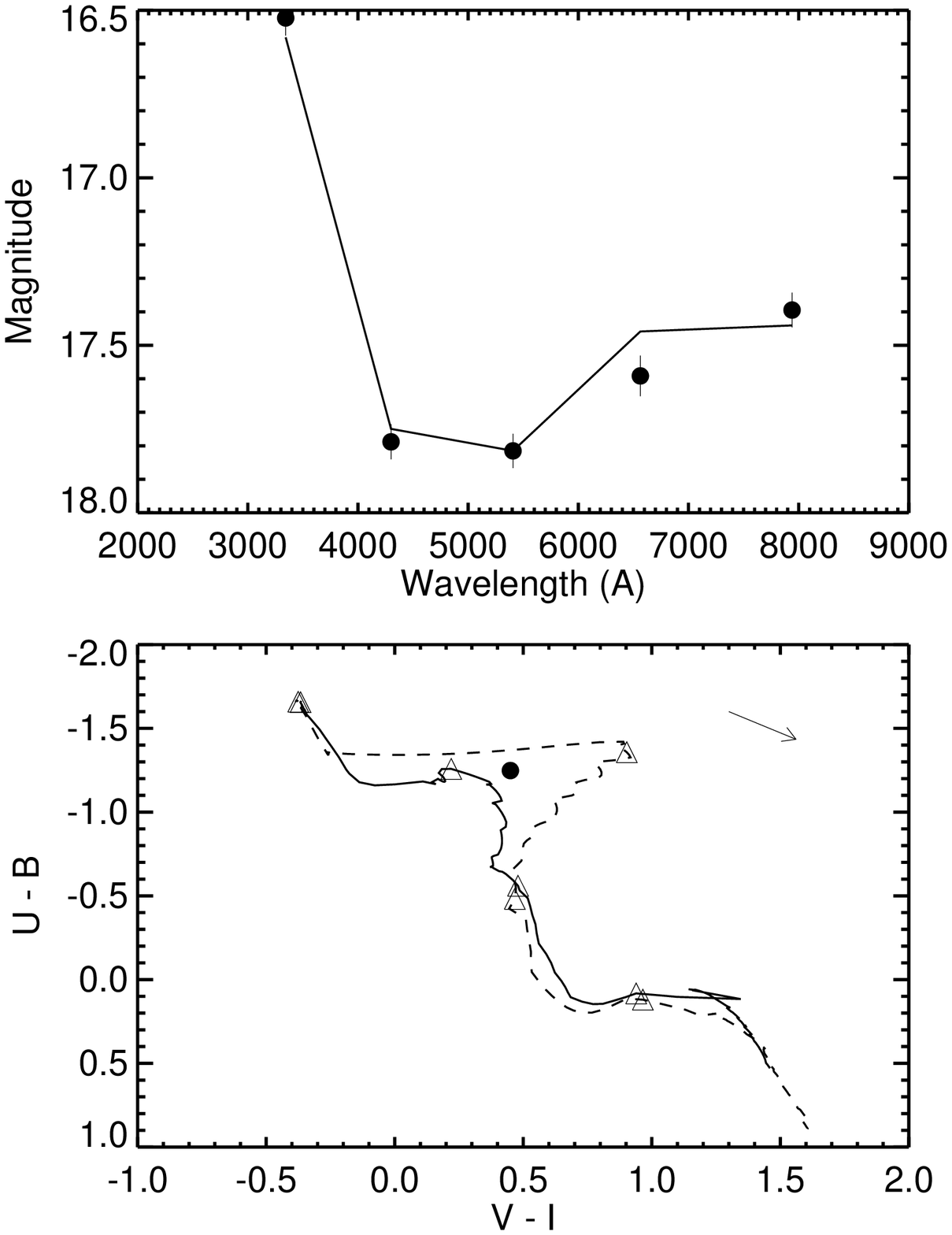}
\caption{Photometric measurements and age determinations for clusters 1 (left) and 10 (right). The top panels show the predicted SEDs from Charlot \& Bruzual models for $Z=0.017$ (2009, private communication), for the age and extinction that best fit the observations (solid line). The data points represent measurements in $U$, $B$, $V$, H$\alpha$, and $I$ filters, and have been corrected for extinction in both the Milky Way and in M83 (values for the latter are listed in Tab.~\ref{tab_reddening}). Note the strong nebular emission in the H$\alpha$ measurement for cluster~1. The bottom panels compare the $U-B$ vs. $V-I$ colors for each cluster with model predictions for metallicities of $Z=0.017$ (solid line) and $Z=0.034$ (dashed line). The arrow shows the direction of reddening. The observed cluster colors have been corrected for reddening in the Milky Way but not in M83. The triangles mark ages of 1, 10, 100, and 1000~Myr along each model, starting from the upper left.}
\label{fig_photsed}
\end{figure}

Our spectroscopic and photometric ages are compiled in columns (1)-(9) of Tab.~\ref{tab_age}, which give (1) the cluster ID; (2) and (4), the spectroscopic ages from the semi-empirical and theoretical models; (3) and (5), the reduced $\chi^2$ value of the best fits to the observed spectra; (6) and (7), the photometric ages from the $Z=0.017$ and $Z=0.034$ models; (8)-(13) ratios of the age estimates derived from the different techniques. The last four rows of Tab.~\ref{tab_age} give the minimum, maximum, mean, and standard deviation values for relevant columns. Figure~\ref{fig_photsed} shows the photometric age determinations for cluster 1 (mean photometric age $\sim$4\,Myr) and cluster 10 (mean photometric age $\sim$19\,Myr). They were derived using the technique described in \cite{cha10}. Figure~\ref{fig_photsed} also shows the measured V-B and V-I colors for clusters 1 and 10. Considering the mean of the spectroscopic and photometric ages, clusters 1 and 10 are representative of young age ($<10$\,Myr) and older age ($>10$\,Myr) clusters, respectively. 

The ages estimated from the four different techniques are in excellent agreement.  Ages derived by comparing UV spectroscopy with semi-empirical and theoretical predictions are within a factor of 1.5.  This is similar to the $1\sigma$ uncertainties of $\sim$1\,Myr found previously for young clusters from UV spectroscopy \citep{tre01,cha03}. Ages determined by comparing the \textit{HST}/WFC3 optical colors of the clusters with predictions from stellar population synthesis models at the two metallicities agree within a factor of 2. Finally, the spectroscopic and photometric ages are in excellent agreement (within a factor of 1.2 on average). Table~\ref{tab_age} shows that most clusters are quite young, with ages $< 5\,$Myr. Clusters 6, 7, and 10, which have the oldest spectroscopic ages, also have the oldest photometric ages. The worst agreement between the spectroscopic and photometric ages is for cluster 8 (difference of a factor of 4). The FUV spectrum of this cluster lacks the strong P-Cygni profiles found in young star clusters, yet the cluster has very blue colors, resulting in a young ($\sim$4\,Myr) photometric age. 

Comparing with previous work, our ages agree with the photometric ages derived by \cite{har01} based on \textit{HST}/WFPC2 photometry, except for clusters 6, 7, and 10, which are older than 6\,Myr in our case. We found 3.6, 3.4, and 4.1\,Myr, for the ages of clusters 1, 2, and 3 respectively (means of spectroscopic and photometric ages), which is in excellent agreement with the ages found by \cite{bre02} for region A (all $\sim$4\,Myr), using the same spectroscopic data as here. On the other hand, \cite{bre02} determined an age of 7\,Myr for their region B, from the equivalent width of H$\beta$. Region B corresponds to the location of our clusters 7-9. We found mean ages of 15.5\,Myr for cluster 7 and 3.4\,Myr for cluster 9. As explained in the previous paragraph, we were unable to pin down the age of cluster 8, which is also in region B. Finally, region D of \cite{bre02} corresponds to a cluster that we did not include in our sample, due to its truncated spectral trace.

We conclude that our spectroscopic and photometric age estimates are within the expected uncertainties for 12 out of 13 of the clusters. The clusters are $\sim$3-20\,Myr old and were not all formed at the same time. In addition, we do not find any compelling evidence for an age gradient along M83's arc-shaped starburst, but rather see clumping of clusters of similar age with very young clusters dominating the northern-most and southern-most portion of the starburst that we studied, and somewhat older clusters found between. This is in disagreement with \cite{pux97} and \cite{dia06}, who have suggested that there is an age gradient along the starburst, as described in the introduction. 

\subsection{Spectroscopic versus Photometric Mass}

Here, we compare the masses estimated for our clusters using spectroscopic and photometric techniques. These are compiled in columns (1)-(7) of Tab.~\ref{tab_mass}, which give (1) the cluster ID; (2) and (3), the masses derived by fitting the semi-empirical and the theoretical models to the FUV spectroscopy; (4) and (5) the masses derived from the optical photometry assuming metallicities of $Z=0.013$ and $Z=0.017$; and (6)-(11) ratios of masses obtained from the different techniques. 

Column (6) of Tab.~\ref{tab_mass} shows that the spectroscopic masses from the semi-empirical models are in very good agreement with those from the theoretical models, and that the semi-empirical masses are systematically larger by a factor of 1.5 on average than the theoretical masses. The latter is because the semi-empirical models have systematically larger values of the mean ratio described in (ii) of the last paragraph of \S~\ref{ana_spec}. Column (7) of Tab.~\ref{tab_mass} shows that the photometric masses are not significantly affected by metallicity. Finally, columns (8) to (11) show that the spectroscopic masses are systematically larger than the photometric masses by a factor of 4-6 on average. We suggest that this is the result of the high background contamination of our wide slit and slitless crowded spectra.

Optical photometry provides more leverage for determining the cluster mass. This is because photometry captures light from low and intermediate mass stars, which contribute more by mass to the IMF than massive stars. In addition, the total light from the cluster is estimated better from the photometry. Therefore, we consider our photometric masses (M$_p$) as more reliable than our spectroscopic masses.

Our most massive cluster has M$_p$ between 2 and $3\times10^5$\,M$_\odot$, comparable to the virial mass of the ionizing cluster of 30 Doradus, NGC 2070 ($4.5\times10^5$\,M$_\odot$, \citealt{bos09}), in the Large Magellanic Cloud. According to \cite{lar10}, young star clusters with masses larger than 10$^5$\,M$_\odot$ can last an age comparable to or exceeding the age of the universe. Therefore, this massive cluster in M83 could be a globular cluster progenitor, while two other clusters, which have M$_p\approx10^5$\,M$_\odot$, may also survive. Cluster 14 has M$_p\approx4\times10^3$\,M$_\odot$ and appears to be the most compact in size in the FUV image. Finally, the rest of the clusters have M$_p$ of a few$\,\times10^4$\,M$_\odot$. 

Note that our older clusters tend to be more massive than the younger ones. This is probably an observational effect. Older clusters at lower masses are harder to detect in the FUV.

\subsection{Intrinsic Reddening}

Columns (1)-(5) of Tab.~\ref{tab_reddening} give (1) the cluster's ID, (2) $\beta_i$, the power-law index of the continuum over the wavelength range $1240-1670\,$\AA, corrected for a foreground Milky Way reddening of $E(B-V)=0.06$; (3) the intrinsic reddening of the cluster derived from the FUV spectra; (4) the intrinsic reddening derived from the optical photometry; and (5) $L_{1500}$, the intrinsic luminosity of the cluster at 1500\,\AA, derived by adopting a distance of 4.5 Mpc to M83.

Overall, the agreement between the reddening values derived from the spectroscopic and the photometric analysis is fair. For clusters 3 and 14, which have dusty local environments (see Fig.~\ref{fig_composite}), the agreement is good. Indeed, both clusters have large intrinsic spectroscopic and photometric reddenings. The agreement is also good for clusters 11 and 12, which appear to be the least reddened by both measures of the intrinsic extinction. However, for cluster 1, E(B-V)=0.18 from the spectroscopy, while E(B-V)$<$0.06 from the photometry. Cluster 3 has the largest intrinsic value of $L_{1500}$, while clusters 4, 13, and 14, which appear faint in the optical images, have the lowest values of $L_{1500}$. 

\subsection{Diffuse Stellar Field}

\begin{figure}
\plotone{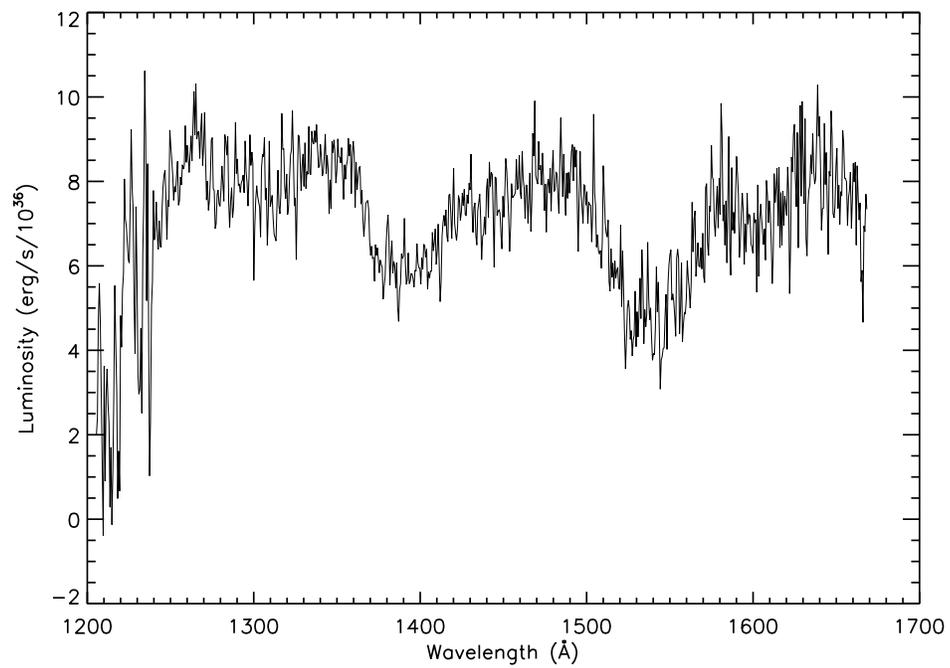}
\caption{Extracted spectrum of the diffuse stellar field. For this spectrum, we performed no background subtraction, reddening correction, or horizontal shifting of the spectral traces.}
\label{fig_fieldsp}
\end{figure}

In addition to compact star clusters, there is UV emission from the diffuse field star population. We obtained the spectrum representing the diffuse stellar field by summing spectra extracted from the long slit exposure, at the locations indicated by the dashed lines in Fig.~\ref{fig_fuvmatch}. We performed no background or reddening correction for the field spectrum. In addition, we did not shift the spectral traces in the dispersion direction, due to the lack of narrow interstellar lines that could be used for reference. Figure~\ref{fig_fieldsp} shows the result. The spectrum lacks the P-Cygni profiles of N V, Si IV, and C IV, which are signatures of O stars, but it shows strong absorption components in these features, which indicates the presence of B stars. This is consistent with a picture where (massive) O stars form mostly in clusters but not in the field, and where the field is dominated by B stars from clusters which disrupt rapidly, on timescales of $\sim$10\,Myr \citep{tre01,cha06,pel07}.

\section{SUMMARY AND CONCLUSIONS}\label{conclusion}

We estimated the ages, reddenings, and masses of 13 clusters within the circumnuclear starburst of M83, the nearest such example, based on \textit{HST}/STIS/FUV-MAMA imaging and spectroscopy, and \textit{HST}/WFC3/UVIS photometry. We derived four ages for each cluster. The first two ages were obtained by fitting the UV spectra with semi-empirical \citep{rob93} and theoretical \citep{lei10} models corresponding to a metallicity of $Z=0.020$, giving the most weight to the spectral features N V 1240, Si IV 1400, and C IV 1500, which are expected to be prominent in the observed spectra of clusters with ages of less than 20\,Myr. Note that theoretical models with $Z=0.040$ yield, on average, ages and masses identical to those at $Z=0.020$. We compared our spectroscopic ages with the photometric ages derived by \cite{cha10} using Charlot and Bruzual stellar population synthesis models corresponding to metallicities of $Z=0.017$ and $Z=0.034$.

The spectroscopic ages derived with the semi-empirical and the theoretical predictions are within a factor of 1.2 on average.  The spectroscopic and photometric ages agree at a similar level. Our ages agree with those derived from \textit{HST}/WFPC2 photometry by \cite{har01}, except for clusters 6, 7, and 10, which are older than 6\,Myr in our case. Our ages for clusters 1-3 agree with the ages of region A derived from STIS FUV spectroscopy by \cite{bre02}. The clusters are $\sim$3-20\,Myr old and were not all formed at the same time. We found no age gradient along M83's starburst, in disagreement with \cite{pux97} and \cite{dia06}. Our clusters with strong P-Cygni profiles have photometric masses of at least a few$\,\times10^4$\,M$_\odot$, seem to have formed stars more massive than $30$\,M$_\odot$, and are consistent with a Kroupa IMF from $0.1-100\,$M$_\odot$. The rest of the clusters are consistent with having formed at least some B stars. B stars dominate the field population. 

\acknowledgments

\begin{figure}
\plotone{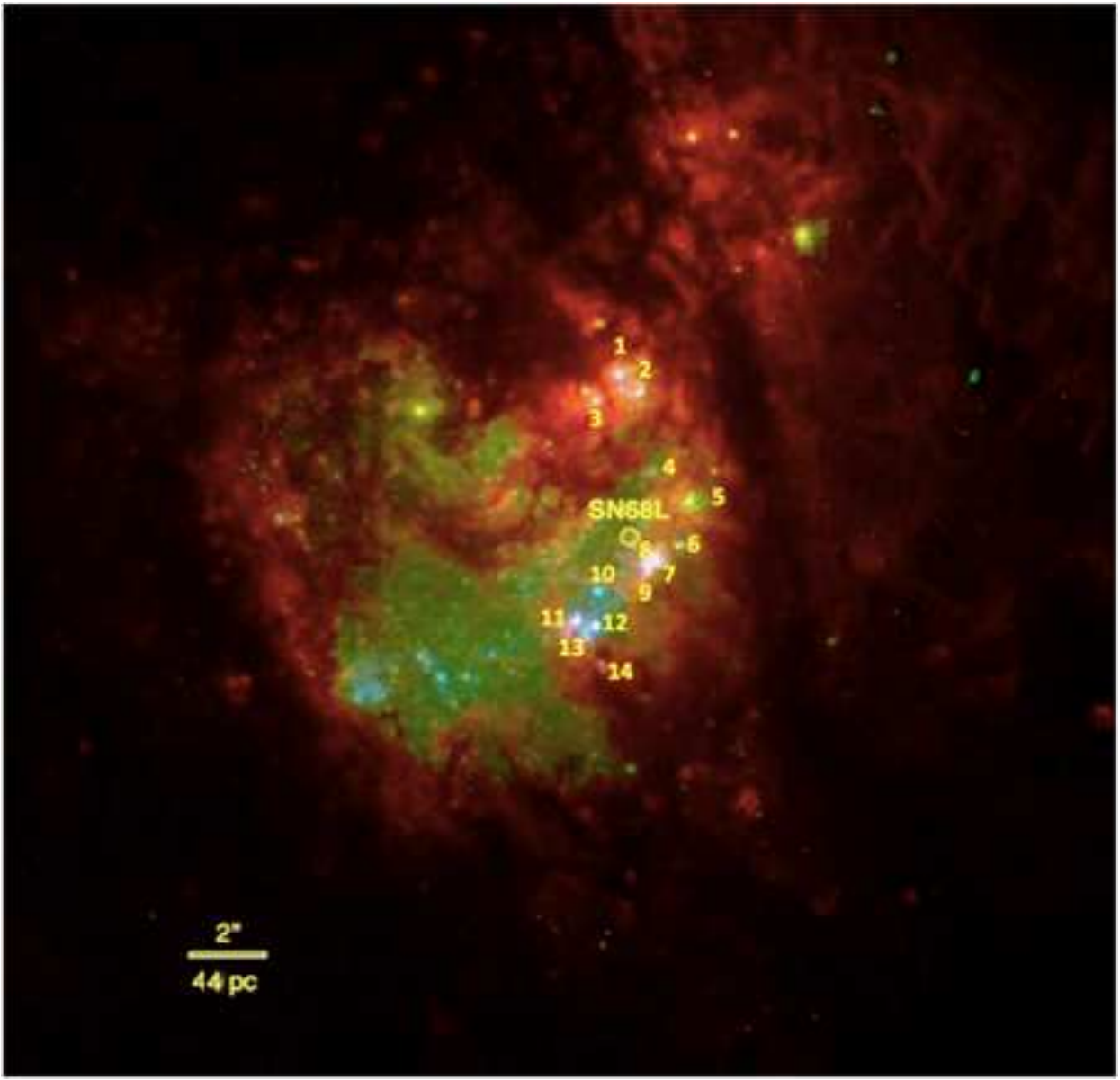}
\caption{\textit{HST}/WFC3/UVIS$+$STIS/FUV-MAMA composite false-color logarithmic-scale image of the two circumnuclear rings of gaseous material embracing M83's starburst. H$\alpha$ is shown in red, the V band in green, and the FUV band in blue. The galaxy's optical nucleus is the large yellow-green dot in the upper left. The starburst clusters in our sample are labeled. North is at the top, and east is to the left. The location of SN1968L is indicated.}
\label{fig_composite}
\end{figure}

This work was supported by NASA grant N1317. R. C. is grateful for support from NSF through CAREER award 0847467. We would like to thank William P. Blair for producing Fig.~\ref{fig_composite}. In addition, we would like to thank Brad Whitmore and our referee for comments that greatly improved the quality of this paper.

{\it Facilities:} \facility{HST (STIS)}, \facility{HST (WFC3)}.

\clearpage

\begin{deluxetable}{lccccccc}
\tablecolumns{7}
\tabletypesize{\scriptsize}
\tablewidth{0pc}
\tablecaption{OBSERVATIONS}
\tablehead{Type & Date & Exp. time  & PA & Aperture & $\Delta\lambda$ & Ext. \\
\hfill & \hfill & (min) & ($^\circ$) & \hfill & (\AA) & \hfill \\
(1) & (2) & (3) & (4) & (5) & (6) & (7)
}
\startdata
slitless spectra & 1998-05-02 & 42 & -24.9 & F25SFR2 & 1275-1700 & sfl \\
slitless spectra & 1998-05-03 & 45 & -24.9 & F25SFR2 & 1275-1700 & sfl \\
slitless spectra & 1998-05-03 & 39 & -24.9 & 25MAMA & 1150-1700 & sfl \\
direct image & 1999-05-19 & 23 & 32.9 & F25QTZ & 1475-1700 & x2d \\
long slit spectra & 2000-04-23 & 42 & -9.5 & 25MAMA & 1150-1700 & flt
\enddata
\label{tab_data}
\end{deluxetable}

\begin{deluxetable}{lccccccccccc}
\tabletypesize{\scriptsize}
\tablecolumns{12}
\tablewidth{0pc}
\tablecaption{OBSERVED PROPERTIES FROM FUV IMAGING AND SPECTROSCOPY}
\tablehead{ID &  $\alpha(J2000)$ & $\delta(J2000)$ &
\multicolumn{3}{c}{ S/N\tablenotemark{a}} &
\multicolumn{3}{c}{$F_{1500}$\tablenotemark{b}} &
\multicolumn{3}{c}{$\beta_{\textrm{obs}}$\tablenotemark{c}} \\
\hfill & (h:m:s) & ($^\circ$:$'$:$''$) &
\multicolumn{3}{c}{\hfill} &
\multicolumn{3}{c}{ (erg s$^{-1}$ cm$^{-s}$ \AA$^{-1}$)} &
\multicolumn{3}{c}{\hfill} \\
(1) & (2) & (3) & 
(4)\tablenotemark{d} & (5)\tablenotemark{e} & (6)\tablenotemark{f}  & 
(7)\tablenotemark{d} & (8)\tablenotemark{e}  & (9)\tablenotemark{f}  & 
(10)\tablenotemark{d} & (11)\tablenotemark{e}  & (12)\tablenotemark{f}
}
\startdata
1 & 13:37:00.49 &  -29:51:54.93 & 18 & 17 & 27 & 6.0E-15 & 8.9E-15 & 8.0E-15 & -1.3 ( -1.1 ) &  -0.7 & -0.5  \\
2 & 13:37:00.45 &  -29:51:55.40 & 16 & 14 & 24 & 4.8E-15 & 6.8E-15 & 6.4E-15 & -1.5 ( -1.2 ) &  -0.9 & -0.9  \\
3 & 13:37:00.54 &  -29:51:55.61 & 14 & 10 & 19 & 3.8E-15 & 4.4E-15 & 4.4E-15 & -0.6 ( -0.4 ) &  0.2 & 0.8  \\
4 & 13:37:00.42 &  -29:51:57.52 & 6 & \nodata  & \nodata  & 9.2E-16 & \nodata  & \nodata  & -0.6 ( -0.7 ) &  \nodata  & \nodata   \\
5 & 13:37:00.34 &  -29:51:58.31 & \nodata  & \nodata  & \nodata  & \nodata  & \nodata  & \nodata  & \nodata     &  \nodata  & \nodata   \\
6 & 13:37:00.37 &  -29:51:59.48 & \nodata & 16 & 23 & \nodata & 5.3E-15 & 5.4E-15 & \nodata    &  -1.4 & -1.5  \\
7 & 13:37:00.42 &  -29:51:59.83 & \nodata & 24 & 35 & \nodata & 1.1E-14 & 1.1E-14 & \nodata    &  -1.4 & -1.5  \\
8 & 13:37:00.44 &  -29:51:59.99 & \nodata & 24 & 36 & \nodata & 1.2E-14 & 1.2E-14 & \nodata    &  -1.6 & -1.6  \\
9 & 13:37:00.43 &  -29:52:00.20 & \nodata & 22 & 33 & \nodata & 9.5E-15 & 1.0E-14 & \nodata    &  -1.6 & -1.7  \\
10 & 13:37:01.00 &  -29:52:00.71 & 21 & 22 & 34 & 7.7E-15 & 1.1E-14 & 1.2E-14 & -1.5 ( -1.3 ) &  -1.4 & -1.3  \\
11 & 13:37:01.00 &  -29:52:01.45 & 34 & 35 & 50 & 2.1E-14 & 2.4E-14 & 2.6E-14 & -2.2 ( -1.9 ) &  -2.1 & -2.2  \\
12 & 13:37:00.54 &  -29:52:01.62 & 34 & 35 & 50 & 2.1E-14 & 2.5E-14 & 2.6E-14 & -2.3 ( -2.0 ) &  -2.1 & -2.2  \\
13 & 13:37:00.56 &  -29:52:01.91 & 22 & 23 & 35 & 9.0E-15 & 1.2E-14 & 1.2E-14 & -2.3 ( -2.0 ) &  -2.0 & -2.1  \\
14 & 13:37:00.53 &  -29:52:02.62 & 7 & \nodata  & \nodata  & 1.5E-15 & \nodata  & \nodata  & -0.7 ( -0.7 ) &  \nodata  & \nodata   
\enddata
\tablenotetext{a}{Median S/N of the extracted one-dimensional STIS spectrum over the common wavelength range $1300-1670\,$\AA.}
\tablenotetext{b}{Reddened flux at 1500\,\AA~measured from STIS spectrum.}
\tablenotetext{c}{Power-law index of reddened continuum ($F\propto\lambda^\beta$) over the wavelength range $1300-1670\,$\AA, which is shared by all spectral exposures. For the long slit spectra we also give the value over the extended range $1240-1690\,$\AA. No data shown if $\beta_{\textrm{obs}}>0$.}
\tablenotetext{d}{From STIS long slit spectrogram taken with $1150-1700\,$\AA~aperture. No data shown for clusters outside of slit.}
\tablenotetext{e}{From STIS slitless spectrogram taken with $1150-1700\,$\AA~aperture. No data shown for clusters with faint spectral traces that overlap.}
\tablenotetext{f}{From summed STIS slitless spectrograms taken with $1275-1700\,$\AA~aperture. No data shown for clusters with faint spectral traces that overlap.}
\label{tab_observed}
\end{deluxetable}

\begin{deluxetable}{lcccccccccccc}
\tabletypesize{\scriptsize}
\tablecolumns{13}
\tablewidth{0pc}
\tablecaption{CLUSTER AGES}
\tablehead{
ID & 
\multicolumn{4}{c}{Spectroscopic\tablenotemark{a}} & 
\multicolumn{2}{c}{Photometric\tablenotemark{b}} & 
\multicolumn{6}{c}{Ratios from Different Techniques}\\
\hfill & 
\multicolumn{4}{c}{(Myr)} & 
\multicolumn{2}{c}{(Myr)} & 
\multicolumn{6}{c}{\hfill} \\
\hfill &
semi\tablenotemark{c} & 
$\chi^2$\tablenotemark{d} & 
theo\tablenotemark{e} & 
$\chi^2$\tablenotemark{d} & 
z017 \tablenotemark{f} & 
z034 \tablenotemark{g} & 
semi/theo & 
z017/z034 & 
semi/z017 & 
theo/z017 & 
semi/z034 & 
theo/z034 \\
(1) & (2) & (3) & (4) & (5) & (6) & (7) & (8) & (9) & (10) & (11) & (12) & (13)
}
\startdata
1	&	3.9	&	3.3	&	2.6	&	5.4	&	4.2	&	3.8	&	1.5	&	1.1	&	0.9	&	0.6	&	1.0	&	0.7	\\
2	&	3.4	&	4.2	&	2.4	&	7.4	&	4.2	&	3.8	&	1.4	&	1.1	&	0.8	&	0.6	&	0.9	&	0.6	\\
3	&	3.5	&	4.8	&	3.5	&	4.0	&	4.0	&	5.5	&	1.0	&	0.7	&	0.9	&	0.9	&	0.6	&	0.6	\\
4	&	3.5	&	1.5	&	2.7	&	1.9	&	4.2	&	5.0	&	1.3	&	0.8	&	0.8	&	0.7	&	0.7	&	0.5	\\
6	&	19.8	&	1.9	&	16.8	&	1.7	&	10.0	&	24.0	&	1.2	&	0.4	&	2.0	&	1.7	&	0.8	&	0.7	\\
7	&	14.2	&	2.3	&	14.3	&	2.9	&	9.5	&	24.0	&	1.0	&	0.4	&	1.5	&	1.5	&	0.6	&	0.6	\\
8	&	15.6	&	2.4	&	14.3	&	2.4	&	4.0	&	3.6	&	1.1	&	1.1	&	3.9	&	3.6	&	4.3	&	3.9	\\
9	&	4.5	&	5.3	&	3.7	&	2.3	&	3.3	&	2.2	&	1.2	&	1.5	&	1.4	&	1.1	&	2.1	&	1.7	\\
10	&	11.8	&	2.9	&	14.1	&	4.7	&	13.2	&	24.0	&	0.8	&	0.5	&	0.9	&	1.1	&	0.5	&	0.6	\\
11	&	3.4	&	6.2	&	2.6	&	8.8	&	6.3	&	4.6	&	1.3	&	1.4	&	0.5	&	0.4	&	0.7	&	0.6	\\
12	&	3.4	&	7.1	&	2.6	&	10.4	&	6.3	&	4.4	&	1.3	&	1.4	&	0.5	&	0.4	&	0.8	&	0.6	\\
13	&	4.2	&	2.6	&	2.8	&	4.9	&	9.1	&	5.0	&	1.5	&	1.8	&	0.5	&	0.3	&	0.8	&	0.6	\\
14	&	2.2	&	2.4	&	2.6	&	4.1	&	3.3	&	2.2	&	0.8	&	1.5	&	0.7	&	0.8	&	1.0	&	1.2   \\
Min	&	2.2	&	1.5	&	2.4	&	1.7	&	3.3	&	2.2	&	0.8	&	0.4	&	0.5	&	0.3	&	0.5	&	0.5	\\
Max	&	19.8	&	7.1	&	16.8	&	10.4	&	13.2	&	24.0	&	1.5	&	1.8	&	3.9	&	3.6	&	4.3	&	3.9	\\
Mean	&	7.2	&	3.6	&	6.5	&	4.7	&	6.3	&	8.6	&	1.2	&	1.1	&	1.2	&	1.0	&	1.1	&	1.0	\\
Stdev	&	5.9	&	1.8	&	5.8	&	2.7	&	3.2	&	8.8	&	0.2	&	0.5	&	0.9	&	0.9	&	1.0	&	0.9
\enddata
\tablenotetext{a}{From STIS long slit ($25"\times2"$) $1205.5-1690$\,\AA~spectroscopy for clusters $1-4$ and $10-14$. From STIS slitless $1205.5-1670$\,\AA~spectroscopy for the rest of clusters. }
\tablenotetext{b}{From WFC3 optical photometry.}
\tablenotetext{c}{Using $Z=0.020$ semi-empirical spectra.} 
\tablenotetext{d}{Reduced $\chi^2$ value of the best model fit to the cluster spectrum.} 
\tablenotetext{e}{Using $Z=0.020$ theoretical spectra.}
\tablenotetext{f}{Using $Z=0.017$ magnitude predictions.}
\tablenotetext{g}{Using $Z=0.034$ magnitude predictions.}
\label{tab_age}
\end{deluxetable}

\begin{deluxetable}{lcccccccccc}
\tabletypesize{\scriptsize}
\tablecolumns{11}
\tablewidth{0pc}
\tablecaption{CLUSTER MASSES}
\tablehead{
ID &
\multicolumn{2}{c}{Spectroscopic\tablenotemark{a}} &
\multicolumn{2}{c}{Photometric\tablenotemark{b}} &
\multicolumn{6}{c}{Ratios from Different Techniques}\\
\hfill & 
\multicolumn{2}{c}{($M_\odot$)} &
\multicolumn{2}{c}{($M_\odot$)} &
\multicolumn{6}{c}{\hfill} \\
\hfill & 
semi\tablenotemark{c} & 
theo\tablenotemark{d} &
z017\tablenotemark{e} & 
z034\tablenotemark{f} & 
semi/theo & 
z017/z034 &
semi/z017 &
theo/z017 &
semi/z034 &
theo/z034 \\
(1) & (2) & (3) & (4) & (5) & (6) & (7) & (8) & (9) & (10) & (11)
}
\startdata
1	&	1.1E+05	&	6.9E+04	&	2.5E+04	&	2.7E+04	&	1.5	&	0.9	&	4.3	&	2.8	&	4.0	&	2.6	\\
2	&	6.0E+04	&	5.2E+04	&	1.9E+04	&	2.1E+04	&	1.2	&	0.9	&	3.1	&	2.7	&	2.9	&	2.5	\\
3	&	1.4E+05	&	1.2E+05	&	4.4E+04	&	4.7E+04	&	1.2	&	0.9	&	3.2	&	2.8	&	3.0	&	2.6	\\
4	&	3.1E+04	&	2.3E+04	&	1.7E+04	&	1.8E+04	&	1.3	&	0.9	&	1.9	&	1.4	&	1.7	&	1.3	\\
6	&	7.4E+05	&	3.8E+05	&	6.4E+04	&	9.9E+04	&	2.0	&	0.6	&	11.6	&	5.9	&	7.5	&	3.8	\\
7	&	1.0E+06	&	6.5E+05	&	1.9E+05	&	3.1E+05	&	1.6	&	0.6	&	5.5	&	3.4	&	3.4	&	2.1	\\
8	&	9.3E+05	&	5.2E+05	&	4.1E+04	&	3.6E+04	&	1.8	&	1.1	&	22.8	&	12.7	&	25.6	&	14.2	\\
9	&	1.2E+05	&	7.7E+04	&	1.8E+04	&	2.0E+04	&	1.5	&	0.9	&	6.4	&	4.2	&	6.0	&	3.9	\\
10	&	5.7E+05	&	4.5E+05	&	6.8E+04	&	1.3E+05	&	1.3	&	0.5	&	8.3	&	6.6	&	4.4	&	3.5	\\
11	&	9.1E+04	&	7.3E+04	&	5.0E+04	&	5.8E+04	&	1.3	&	0.9	&	1.8	&	1.4	&	1.6	&	1.2	\\
12	&	7.9E+04	&	6.3E+04	&	5.5E+04	&	4.8E+04	&	1.3	&	1.1	&	1.4	&	1.1	&	1.6	&	1.3	\\
13	&	4.3E+04	&	2.3E+04	&	2.6E+04	&	1.6E+04	&	1.8	&	1.6	&	1.7	&	0.9	&	2.7	&	1.4	\\
14	&	3.0E+04	&	2.8E+04	&	3.9E+03	&	4.2E+03	&	1.1	&	0.9	&	7.6	&	7.2	&	7.1	&	6.7	\\
Min	&	3.1E+04	&	2.3E+04	&	1.7E+04	&	1.6E+04	&	1.2	&	0.5	&	1.4	&	0.9	&	1.6	&	1.2	\\
Max	&	1.0E+06	&	6.5E+05	&	1.9E+05	&	3.1E+05	&	2.0	&	1.6	&	22.8	&	12.7	&	25.6	&	14.2	\\
Mean	&	3.3E+05	&	2.1E+05	&	5.1E+04	&	6.9E+04	&	1.5	&	0.9	&	6.1	&	4.1	&	5.5	&	3.6	\\
Stdev	&	3.8E+05	&	2.2E+05	&	4.7E+04	&	8.3E+04	&	0.3	&	0.3	&	5.9	&	3.3	&	6.4	&	3.5	\\
\enddata
\tablenotetext{a}{From STIS long slit ($25"\times2"$) $1205.5-1690$\,\AA~spectroscopyfor clusters $1-4$ and $10-14$. From STIS slitless $1205.5-1670$\,\AA~spectroscopy for the rest of clusters.} 
\tablenotetext{b}{From WFC3 optical photometry.}
\tablenotetext{c}{Using $Z=0.020$ semi-empirical spectra.}
\tablenotetext{d}{Using $Z=0.020$ theoretical spectra.}
\tablenotetext{e}{Using $Z=0.017$ magnitude predictions.}
\tablenotetext{f}{Using $Z=0.034$ magnitude predictions.}
\label{tab_mass}
\end{deluxetable}

\begin{deluxetable}{lccccc} \tabletypesize{\scriptsize}
\tablecolumns{6}
\tablewidth{0pc}
\tablecaption{INTRINSIC REDDENINGS}
\tablehead{
ID & $\beta_i$\tablenotemark{a} & $E(B-V)_i$ & $E(B-V)_i$ & $E(B-V)_i$ & $L_{1500}$\tablenotemark{e} \\
\hfill & \hfill & spec\tablenotemark{b} & z017\tablenotemark{c}  & z034\tablenotemark{d} & (erg s$^{-1}$ \AA$^{-1}$) \\
(1) & (2) & (3) & (4) & (5) & (6)
}
\startdata
1	&	-1.7	&	0.18	&	0.02	&	0.06	&	1.4E+38	\\	
2	&	-1.9	&	0.16	&	0.06	&	0.10	&	9.4E+37	\\	
3	&	-0.9	&	0.28	&	0.28	&	0.20	&	2.0E+38	\\	
4	&	-1.0	&	0.23	&	0.24	&	0.20	&	5.0E+37	\\	
6	&	-1.8	&	0.13	&	0.20	&	0.04	&	8.8E+37	\\	
7	&	-1.7	&	0.13	&	0.22	&	0.06	&	1.9E+38	\\	
8	&	-1.9	&	0.10	&	0.16	&	0.16	&	1.5E+38	\\	
9	&	-1.9	&	0.10	&	0.14	&	0.18	&	1.3E+38	\\	
10	&	-1.9	&	0.14	&	0.02	&	0.00	&	1.3E+38	\\	
11	&	-2.5	&	0.05	&	0.00	&	0.08	&	1.5E+38	\\	
12	&	-2.6	&	0.04	&	0.00	&	0.00	&	1.3E+38	\\	
13	&	-2.6	&	0.03	&	0.00	&	0.02	&	4.8E+37	\\	
14	&	-1.1	&	0.23	&	0.18	&	0.22	&	5.7E+37	
\enddata
\tablenotetext{a}{Power-law index of continuum ($F\propto\lambda^\beta$) corrected for a foreground Milky Way reddening of $E(B-V)=0.06$, over range $1300-1670$\,\AA~(slitless exposure) or $1300-1690$\,\AA~(long slit exposure).}
\tablenotetext{b}{Spectroscopic reddening derived adopting the starburst obscuration law of \cite{cal00} and assuming an intrinsic cluster slope of $\beta=-2.6$.}
\tablenotetext{c}{Photometric reddening using color predictions for $Z=0.017$.}
\tablenotetext{d}{Photometric reddening using color predictions for $Z=0.034$.}
\tablenotetext{e}{Reddening-corrected luminosity at 1500\,\AA~derived from the spectrum of each cluster by adopting the spectroscopically derived value of $E(B-V)_i$ and a distance of $4.5\,$Mpc to M83.}
\label{tab_reddening}
\end{deluxetable}

\end{document}